\documentclass[journal,twoside,web]{ieeecolor}
\usepackage{bbold}
\usepackage{generic}
\usepackage{cite}
\usepackage{amsmath,amssymb,amsfonts}
\usepackage{algorithm}
\usepackage{algorithmic}
\usepackage{graphicx}
\usepackage{textcomp}
\usepackage{ulem}
\usepackage{bm}
\usepackage{xcolor}
{}
\newtheorem{theorem}{Theorem}{}
\newtheorem{lemma}{Lemma}{}
\newtheorem{assumption}{Assumption}
\newtheorem{corollary}{Corollary}{}
{}
\newtheorem{remark}{Remark}{}

\def\BibTeX{{\rm B\kern-.05em{\sc i\kern-.025em b}\kern-.08em
    T\kern-.1667em\lower.7ex\hbox{E}\kern-.125emX}}
\begin{document}

%	\definecolor{backgroundcolor}{RGB}{199, 238, 206}
%	\pagecolor{backgroundcolor}
\title{Observation of Periodic Systems: Bridge Centralized Kalman Filtering and Consensus-Based Distributed Filtering}
\author{Jiachen Qian, Zhisheng Duan*, \IEEEmembership{Senior member, IEEE}, Peihu Duan, Zhongkui Li, \IEEEmembership{Senior member, IEEE}
\thanks{This work is supported by the National Natural Science Foundation of China under Grants T2121002, 62173006. (*$Corresponding\;author: Zhisheng\;Duan)$}
\thanks{J. Qian, Z. Duan and Z. Li are with State Key Laboratory for Turbulence and Complex Systems, Department of Mechanics and Engineering Science, College of
Engineering, Peking University, Beijing, 100871, China. }
\thanks{P. Duan is with the Department of Electronic and Computer Engineering, the Hong Kong University of Science and Technology, Clear Water Bay, Kowloon, Hong Kong, China. E-mails: duanzs@pku.edu.cn (Z. Duan), 1901111654@pku.edu.cn (J. Qian), duanpeihu.work@gmail.com(P. Duan), zhongkli@pku.edu.cn (Z. Li)}}

\maketitle

\begin{abstract}
	 Compared with linear time invariant systems, linear periodic system can describe the periodic processes arising from nature and engineering more precisely. However, the time-varying system parameters increase the difficulty of the research on periodic system, such as stabilization and observation. This paper aims to consider the observation problem of periodic systems by bridging two fundamental filtering algorithms for periodic systems with a sensor network: consensus-on-measurement-based distributed filtering (CMDF) and centralized Kalman filtering (CKF).
	  Firstly, one mild convergence condition based on uniformly collective observability is established for CMDF, under which the filtering performance of CMDF can be formulated as a symmetric periodic positive semidefinite (SPPS) solution to a discrete-time periodic Lyapunov equation. Then, the closed form of the performance gap between CMDF and CKF is presented in terms of the information fusion steps and the consensus weights of the network. Moreover, it is pointed out that the estimation error covariance of CMDF exponentially converges to the centralized one with the fusion steps tending to infinity. Altogether, these new results establish a concise and specific relationship between distributed and centralized filterings, and formulate the trade-off between the communication cost and distributed filtering performance on periodic systems. Finally, the theoretical results are verified with numerical experiments.

\end{abstract}


\section{Introduction}
Periodic systems, as an intermediate class between the time-invariant systems and time-varying systems, take an important part in control theory and engineering. As stated in \cite{bittani1988}, periodic processes arise in nature and engineering behaviors spontaneously or elaborately. Specifically, a wide range of practical systems, such as orbiting satellites with nonlinear oscillation \cite{cole1992nonlinear}, biological systems with periodically instant modulated impulses \cite{churilov2012state}, hard disk drives with irregular sampling rates \cite{nie2011optimal}, and periodically scheduled multirate systems \cite{meyer1975unified, shi2011optimal}, can all be modeled as time-varying periodic systems. Therefore, the study of periodic systems, such as stabilization and observation, can efficiently facillitate the corresponding engineering practice.  

In the past several decades, observation problems of periodic systems have been extensively studied \cite{bittani1988,xie1991h,soderstrom2005periodic,shi2011optimal,dragan2012optimal,shi2013approximate,orihuela2014periodicity}. To name a few, Bittanti {\it et al.} \cite{bittani1988} analyzed the performance of the centralized Kalman filter on periodic systems and proposed a convergence condition based on the symmetric periodic positive semidefinite (SPPS) solution to the corresponding discrete-time periodic Riccati equation (DPRE). Xie and de Souza \cite{xie1991h} considered the $H_{\infty}$ filtering problem of periodic systems with parameter uncertainties and formulated a parameter design method in terms of solving two periodic Riccati differential equations. Shi {\it et al.} \cite{shi2011optimal} proposed a class of optimal periodic sensor schedule policies to minimize the state estimation error for linear time invariant (LTI) systems and proposed the closed-form expression of the optimal performance of the corresponding periodic scheduled filter. 

%The study on the observation problems of periodic systems is sufficient, while the literature of distributed filtering on periodic systems is relatively rare, compared with the aforementioned centralized case. 
However, the aforementioned estimators strongly depend on a central authority for designing and performing, which is fragile to node failure and malicious attack. To overcome these disadvantages,
%enhance the resilience and reduce the communication cost of estimators, 
various distributed frameworks with a sensor network have been proposed and studied \cite{olfati2007distributed,olfati2009kalman,cattivelli2010diffusion,kamal2013information,battistelli2014kullback,Battistelli2015Linear,Wang20181300,he2020distributed,duan2020distributed,li2020Bound,Sayed20219353995,Wu201863,chenb9416784,zhang2016event,shen2019distributed,wanrb8401335,li2017distributed, lijun8651311,LiuRoundRobin8605375}, where each sensor acts as a local fusion node that collects information from its neighbors to reconstruct the state {of the whole system}. For LTI systems, the existing literature of distributed filtering mainly focused on consensus-based distributed filtering algorithms \cite{olfati2007distributed,olfati2009kalman,cattivelli2010diffusion,kamal2013information,battistelli2014kullback,Battistelli2015Linear,Wang20181300,he2020distributed,duan2020distributed,li2020Bound,Sayed20219353995}, where each sensor equipped with the average consensus technique first attempted to collect the neighbouring information and then altogether guaranteed the stability of the estimator under weak local observability. While for linear periodic systems, the existing literature of distributed filtering mainly focused on the stability analysis, or sub-optimality analysis with restricted communication resources \cite{zhang2016event,shen2019distributed,wanrb8401335,li2017distributed, lijun8651311,LiuRoundRobin8605375,Wu201863,chenb9416784}. Specifically, Li {\it et al.} \cite{li2017distributed, lijun8651311} proposed a linear matrix inequalities (LMIs)-based parameter design technique to compensate for the effect of packet loss and time delay for distributed state estimation of periodic systems. Liu {\it et al.} \cite{LiuRoundRobin8605375} raised a Round-Robin-protocal-based time-varying distributed filter for multirate systems to simultaneously reduce the communication among sensor nodes and guarantee the stability of the filter. Wu {\it et al.} \cite{Wu201863} and Chen {\it et al.} \cite{chenb9416784} applied distributed methods to general observation problem of periodic systems such as temperature monitoring. 

One inherent common drawback of existing literature of distributed filtering on periodic systems \cite{zhang2016event,shen2019distributed,wanrb8401335,li2017distributed, lijun8651311,LiuRoundRobin8605375} is that the parameter design procedure essentially depends on the solutions to complex LMIs, and the stability of the filter is directedly related to the solvability of these LMIs. Therefore, there does not exist an intuitional and neat relationship between the stability condition of the distribtued filter and the observability of the periodic system. In addition, the performance evaluation of the aforementioned distributed filters is relatively hard and the trade-off between the communication cost and estimation performance is correspondingly challenging to establish.
Note that classical consensus-based distributed filtering techniques \cite{olfati2007distributed, olfati2009kalman, kamal2013information, talebi2019distributed, liCMweight} have near-optimal estimation performance and concise parameter tuning procedure, when applied for LTI systems.  
%fusion based distribtued filtering technique \cite{battistelli2014kullback, battistelli2016stability, duan2020distributed, he2018consistent, cattivelli2010diffusion}, 
However, the theoretical foundation of applying these consensus-based techniques to periodic systems is still missing.
%which is due to the difficulties induced by time-variant property of the system and the complex correlation of the estimation error between sensor nodes. 
Motivated by above discussions, this paper aims to establish the performance analysis of consensus-based distributed filtering algorithm on periodic linear systems, and to provide guidelines for the design of filtering parameters such as fusion step and fusion weights to optimize the performance.

 \textcolor{black}{Previous exploration of consensus-on-measurement-based distributed filtering (CMDF) algorithm reveals that for LTI systems, the steady-state error covariance matrix can be simplified as the solution to a modified discrete-time Lyapunov equation. In addition, with the number of fusion step $L$ between two successive sampling instants tending to infinity, the performance of the consensus-based distributed filtering algorithm exponentially converges to the centralized optimal case \cite{qian2021consensusbased}.} \textcolor{black}{These theoretical results for LTI systems inspire us to depict the performace of CMDF on periodic systems.  
 Hence, two essential issues of the performance analysis of CMDF induced by the periodicity of the systems are discussed in detail.
 The first one is to formulate the asymptotic periodicity of the error covariance matrix under the uniform observability condition. The other is to bridge the CKF and CMDF, with a closed-form expression of the periodic performance gap compared with the centralized optimal SPPS performance proposed in \cite{bittani1988}, especially interpreting the effect of fusion step $L$.}

The main contributions of this paper are summarized as follows:

\begin{enumerate}
	\item 
	The closed-form SPPS performance of the CMDF algorithm is formulated.
	The asymptotically periodic error covariance matrix of the {CMDF} algorithm on periodic system with a finite fusion step $L$ is derived ({\bf Theorem \ref{thmconverge}}). {It is shown} that as long as the pair of the state transition matrix and the modified observation matrix for each sensor is uniformly collectively observable, the SPPS performance of the estimation error covariance matrix can be simplified as the solution to a {discrete-time periodic Lyapunov equation (DPLE)}. This result quantitatively describes the SPPS performance degradation resulted by insufficient information fusion of CMDF on periodic systems. 
%	{which is an extension} to the literature for LTI system \cite{olfati2007distributed,Kamagarpour4738989,li2020Bound,qian2021consensusbased}. 
	
	\item {The relationship between the fusion step $L$ and the SPPS performance of CMDF algorithm is established for periodic systems. 
%	Based on the formulated infinite series expression of the difference between two DPRE solutions, 
	{It is proved} that the gap between the centralized optimal SPPS performance and the distributed SPPS case exponentially convergents to zero, with $L$ tending to infinity ({\bf Theorems \ref{thmric},\ref{thmcov}}). In addition, the upper bound of the decay rate of the performance gap can be arbitrarily close to the norm of the second largest eigenvalue of the doubly-stochastic weighting matrix of the communication graph ({\bf Corollary \ref{exponential}}).
	This result theoretically reveals the trade-off between the estimation accuracy and the communication cost {in the distributed setting}, and connects the filtering performance with the spectral property of communication graph, which was rarely discussed in the classical distributed filtering literature \cite{olfati2007distributed,kamal2013information,battistelli2014kullback}.} 

	\item 
	Some new properties of the DPLE and DPRE are derived, 
	including the asymptotic convergence of the periodic Lyapunov iteration ({\bf Lemma \ref{lmlya}}) and the uniform boundedness of the norm of feedback {\it monodromy matrices} for a group of DPRE ({\bf Lemma \ref{lmstable}}). {These novel properties {play an essential role} in facilitating the applicability of the CMDF algorithm for periodic systems}.
\end{enumerate}    

The remainder of this paper is organized as follows. {Some preliminaries}, including the problem formulation and some useful lemmas, are presented in Section \ref{sec2}. The main results, including the derivation and analysis of the SPPS performance of the CMDF algorithm on periodic systems, are presented in Sections \ref{sec3} and \ref{bridge}. Some illustrative numerical simulations are presented in Section \ref{sec4}. Conclusions are {drawn} in Section \ref{sec5}.

\textit{Notation:} For two symmetric matrices $X_{1}$ and $X_{2}$, $X_{1}> X_{2}\left(X_{1}\ge X_{2}\right)$ means that $X_{1}-X_{2}$ is positive definite (positive semi-definite). {$exp\left(\cdot\right)$} denotes the exponential function. $\big|a\big|$ denotes the absolute value of real number $a$ or the norm of complex number $a$.  $\mathcal{L}\rhd 0\left(\unrhd 0\right)$ means that all the elements of matrix $\mathcal{L}$ are positive (non-negative). $\mathbb{E}\left\lbrace x\right\rbrace$ denotes the expectation of a random variable $x$. $\lambda\left(A\right)$ denotes the eigenvalue of matrix $A$. $\rho\left(A\right)$ denotes the spectral radius of $A$.  $\left\|A\right\|_2$ denotes the 2-norm (the largest singular value) of matrix $A$.

\section{Preliminaries and Problem Formulation}\label{sec2}
% In this paper, to realize the average consensus, we need to require the $\mathcal{L}$ to be doubly-stochastic and symmetric, which is similar to \cite{yi8678811,Xiao200465}.
\subsection{{System} Model}
{Consider} a network of $N$ sensors, which measure and estimate the states of a target system, described as
\begin{equation}
\begin{aligned}
&x_{k+1}=A_k x_k + \omega_k,\quad k = 0,1,2,\ldots,\\
&y_{i,k}=C_{i,k}x_k + v_{i,k},\quad i = 1,2,\ldots,N,
\end{aligned}
\end{equation}
where $x_k\in\mathbb{R}^n$ is the state vector of the system, $y_{i,k}\in\mathbb{R}^{n_i}$ is the measurement vector of sensor $i$,  $\omega_k\in\mathbb{R}^n$ is the process noise with covariance matrix $Q_k\in\mathbb{R}^{n\times n}$, and $v_{i,k}\in\mathbb{R}^{n_i}$ is the observation noise with covariance matrix  $R_{i,k}\in\mathbb{R}^{n_i \times n_i}$. The sequences $\left\{\omega_k\right\}^{\infty}_{k=0}$ and $\left\{v_{i,k}\right\}^{\infty,N}_{k=0,i=1}$ are assumed to be \textcolor{black}{mutually uncorrelated white Gaussian noise}. $Q_k$ and $R_{i,k}$ are positive definite. Besides, $A_k$ is the state-transition matrix and $C_{i,k}$ is the observation matrix of sensor $i$. $C_k=\left[C_{1,k}^T,C_{2,k}^T,\dots,C_{N,k}^T\right]^T$ and $R_k=\mathrm{diag}\left\{R_{1},\dots,R_{N} \right\}$ are the observation matrix and the covariance matrix of observation noise of the whole network, respectively. Denote $m\triangleq\sum_{i=1}^{N}n_i$ as the dimension of the observation for the whole network. $A_k,\,C_k,\,Q_k,\,R_k$ are $\mathcal{T}$-periodic matrices, i.e., $A_{k+\mathcal{T}}=A_k,\;C_{k+\mathcal{T}}=C_k,\;Q_{k+\mathcal{T}}=Q_k,\;R_{k+\mathcal{T}}=R_k$ for all $k\in \mathbb{N}$. Similar to the notations used in \cite{bittani1988}, $A_.,C_.,Q_.,R_.$ are used in this paper to represent $A_k,C_k,Q_k,R_k,\;\forall k\in\mathbb{N}$, respectively.

The communication topology of the sensor network is denoted by a graph $\mathcal{G}=\left(\mathcal{V},\mathcal{E},\mathcal{L}\right)$, where $\mathcal{V}=\left\lbrace1,2,\dots,N\right\rbrace$ is the node set, $\mathcal{E}\subseteq \mathcal{V}\times\mathcal{V}$ is the edge set, and $\mathcal{L}=\left[l_{ij}\right]$ is the adjacency matrix of the network. The adjacency matrix reflects the interactions among the nodes, $l_{ij}>0\Leftrightarrow\left(i,j\right)\in\mathcal{E}$, which means that sensor $i$ can receive the information from sensor $j$. In this case, sensor $j$ is called an in-neighbor of sensor $i$, and sensor $i$ is called an out-neighbor of sensor $j$. {For simplicity, let $i\in\mathcal{V}$ represent the $i$-th sensor of the network.} $\mathcal{N}_{i}$ denotes the in-neighbor set of sensor $i$, and $l_{i}$ denotes the $i$-th row of $\mathcal{L}$, which contains the information of $\mathcal{N}_{i}$. \textcolor{black}{In addition, the matrix $\mathcal{L}$ is doubly stochastic, i.e, $\sum_{j=1}^{N}l_{ij}=\sum_{j=1}^{N}l_{ji}=1,\forall i\in\mathcal{V}$, {which ensures the average weighting} of the information from the neighboring nodes}. $l_{ij}^{(L)}$ is the $(i,j)$-th element of matrix $\mathcal{L}^L$,  {which is the network adjacency matrix at step $L$.} {The diameter $d$ of graph $\mathcal{G}$ is the length of the longest path between two nodes in the graph.} 

\subsection{Revisit Consensus-Based Distributed Filtering}
{In this subsection, necessary introductions of consensus-based distributed filtering algorithms are provided.
To the best of the authors' knowledge, up to now, consensus-based distributed filtering algorithm are mainly applied for observing LTI systems rather than periodic systems.  
To name a few approaches, Olfati-Saber \cite{olfati2007distributed} {first formulated a framework of consensus-on-measurement-based distributed filtering (CMDF)}, where each sensor obtains the average of the global measurement information at every sampling instant through sufficient fusion with neighbors. Kamal {\it et al.} \cite{kamal2013information} and Battistelli {\it et al.} \cite{battistelli2014kullback,Battistelli2015Linear} proposed some distributed information fusion techniques such as consensus on information matrix, hybrid consensus on measurement and information matrix, where the applied covariance intersection operation can guarantee the stability of the filter with a low fusion step $L$. Duan {\it et al.} \cite{duan229750909} proposed CMDF technique for continuous systems with time-correlated noise and gave the asymptotic optimality analysis. }

\textcolor{black}{For consensus-on-information-based distributed filtering algorithm (CIDF) proposed in \cite{battistelli2014kullback,Battistelli2015Linear}, due to the reduced weights of the observations, the performance gap compared with the centralized optimal case may be sufficiently large when the fusion step $L$ is large. The CMDF algorithm, on the other hand, can achieve the centralized optimal filtering performance to any degree of accuracy, i.e., the performance gap between the centralized and the CMDF algorithm can be arbitrarily small as the fusion step $L$ becomes sufficiently large \cite{qian2021consensusbased}.
Hence, this paper intends to provide theoretical analysis for CMDF algorithm on linear periodic systems to bridge centralized Kalman filtering and consensus-based distributed filtering.}

\subsection{Problem Formulation}
In this subsection, the CMDF algorithm for periodic system is firstly proposed.  
This algorithm is shown in Algorithm \ref{alg1}, 
% \textcolor{blue}{where $N$ is the {\it a priori} parameter shared by all the nodes in the sensor network.} 
{where $N$ is the {\it a priori} parameter shared by all nodes in the sensor network, and the doubly stochastic matrix $\mathcal{L}$ can be shared as global parameters or obatined in a distributed way when the communication graph is strongly connected \cite{GHARESIFARD2012539}.} 
\begin{algorithm}
	\caption{Distributed Information Fusion Algorithm}
	\label{alg1}
	\textbf{Input:}\\
	$\hat{x}_{i,0|0}, P_{i,0|0}\quad i=1,2,\dots,N$\\
	\textbf{Prediction:}\\
	$\hat{x}_{i,k|k-1}=A_{k-1}\hat{x}_{i,k-1|k-1}\\ P_{i,k|k-1}=A_{k-1}P_{i,k-1|k-1}A_{k-1}^T+Q_{k-1}$\\
	\textbf{Fusion:}\\
	Set 
	$$
	S_{i,k}^{(0)}=NC_{i,k}^TR_{i,k}^{-1}C_{i,k}, \qquad I_{i,k}^{(0)}=NC_{i,k}^TR_{i,k}^{-1}y_{i,k}
	$$\\
	For $h=1,2,\dots,L$
	$$
	\begin{aligned}
	S_{i,k}^{(h)}=\sum_{j=1}^{N}l_{ij}S_{j,k}^{(h-1)}, \qquad I_{i,k}^{(h)}=\sum_{j=1}^{N}l_{ij}I_{j,k}^{(h-1)}\;\;
	\end{aligned}
	$$
	\textbf{Correction:}\\
	$P_{i,k|k}=\big(P_{i,k|k-1}^{-1}+S_{i,k}^{(L)}\big)^{-1}$\\
	$\hat{x}_{i,k|k}=P_{i,k|k}\big(P_{i,k|k-1}^{-1}\hat{x}_{i,k|k-1}+I_{i,k}^{(L)}\big)$\\
\end{algorithm}

Note that the parameter $L$, i.e., the fusion step between two successive sampling instants, plays an important part in the structure of the consensus-based algorithm. \textcolor{black}{In \cite{olfati2007distributed,Kamagarpour4738989,kamal2013information}, the fusion step was required to be infinite to obtain the average information of the whole sensor network such that each sensor can obtain the centralized optimal performance. In \cite{battistelli2014kullback,talebi2019distributed,BATTILOTTI2021109589}, the performance of consensus-based algorithms with finite fusion step $L$ is discussed but the closed form of the relationship between performance gap and fusion step $L$ was not formulated. Moreover, the periodicty of the system parameter also increases the difficulty of performance evaluation.}
Hence, {in this paper}, the fundamental problem of CMDF with finite fusion step $L$ for the discrete-time periodic systems is formulated {as follows:}
\begin{enumerate}
	\item {Evaluate} the effect induced by insufficient fusion on the performance of the CM-based distributed filtering algorithm on periodic systems.
	\item {Find out} the closed-form relationship between fusion step $L$ and performance degradation compared with the centralized optimal SPPS performance.
\end{enumerate}

\subsection{Useful Lemmas}
\textcolor{black}{\begin{lemma}\label{lm1}
		\cite{qian2021consensusbased} If the communication topology corresponding to $\mathcal{L}$ is strongly connected and $\mathcal{L}$ is doubly stochastic, then the matrix $\mathcal{L}^k$ converges to $\frac{1}{N}\textbf{1}\textbf{1}^T$ exponentially as $k\to\infty$, i.e., there exist two parameter $M>0,\;0<q<1$, such that
		$$
		\big\|\mathcal{L}^k-\frac{1}{N}\textbf{1}\textbf{1}^T\big\|_2\leq Mq^k.
		$$
		In addition, the parameter $q$ can be chosen as close to the norm of the second largest eigenvalue of $\mathcal{L}$ as possible.
	\end{lemma}}
%	\begin{proof}
%		See the proof of Lemma 1 in .
%		%	The strongly connected property guarantees that the matrix $\mathcal{L}$ is irreducible. With Theorem 8.5.2 and Lemma 8.5.5 in \cite{Horn1985}, the matrix $\mathcal{L}$ is also primitive with positive diagonal elements. Thus, there is only one eigenvalue of $\mathcal{L}$ equal to the spectral radius 1, and the norms of all other eigenvalues are strictly less than one. Consider any eigenvector corresponding to an eigenvalue $\big|\lambda\big|<1$, denoted as ${\it\bf x}$. Then, one has
%		%	$$
%		%	{\bf 1}^T\mathcal{L}{\bf x}={\bf 1}^T{\bf x}=\lambda\textbf{1}^T\textbf{x},
%		%	$$
%		%	where the first equality follows from the fact that $\mathcal{L}$ is doubly stochastic. Therefore one has $\textbf{1}^T\textbf{x}=0$. Thus, all the eigenvalues of $\mathcal{L}$ except 1 are also eigenvalues of $\mathcal{L}-\frac{1}{N}\textbf{1}\textbf{1}^T$. With Theorem 1 in \cite{Xiao200465}, the matrix $\mathcal{L}^k$ converges to $\frac{1}{N}\textbf{1}\textbf{1}^T$ exponentially with the increase of $k$, and the convergence rate is not slower than the norm of the second largest eigenvalue of $\mathcal{L}$.
%    \end{proof}
\vspace{6pt}
\begin{lemma}\label{lm2}
	(Matrix {Inversion} Lemma) \cite{guttman1946enlargement}
	For any matrix $P,Q,C$ of proper dimensions, if $P^{-1}$ and $Q^{-1}$ exist, then the following {equality} holds:
	$$
	\left(P^{-1}+C^TQ^{-1}C\right)^{-1}=P - PC^T\left(CPC^T+Q\right)^{-1}CP.
	$$
\end{lemma}
\vspace{6pt}
\begin{lemma}\label{lm3}
	{\cite{qian2021consensusbased} For any matrix $A \in \mathbb{R}^{n \times n}$}, the following inequality holds: 
	$$
	\left\|A^k\right\|_2\leq\sqrt{n}\sum_{j=0}^{n-1}\binom{n-1}{j}\binom{k}{j}\big\|A\big\|_2^j\rho\left(A\right)^{k-j},
	$$ 
	\textcolor{black}{where $\binom{m}{n}$ is the combinatorial number, $\binom{k}{j}=0$ for $j>k$, $\rho\left(A\right)$ and $\big\|A\big\|_2$ denote the spectral radius and maximum singular value of $A$, respectively}.
\end{lemma}

\section{Convergence Analysis}\label{sec3}
\subsection{Basic Assumptions}
In this section, the convergence property of the proposed CMDF algorithm for periodic system will be analyzed, including the convergence of the parameter matrix and the error covariance matrix to the SPPS solutions of DPRE and DPLE, respectively.  {To do so, the following two assumptions are needed}.
\vspace{6pt}
\begin{assumption}\label{as1}
	The communication topology is strongly connected.
\end{assumption}
\vspace{6pt}
\begin{assumption}\label{as2}
	\cite{bittani1988} The $\mathcal{T}$-periodic system pair $\left(A_{.},C_{.}\right)$ is uniformly observable, i.e., there exists a $\mathcal{T}$-periodic matrix sequence $L_.$, with $L_k:\;\mathbb{Z}\to \mathbb{R}^{n\times m}$, such that the feedback system $A_{.}+L_{.}C_{.}$ is asymptotically stable, i.e., $\prod_{k=1}^{\mathcal{T}}(A_{k}+L_{k}C_{k})$ is Schur stable.
\end{assumption}

\vspace{6pt}
\textcolor{black}{
\begin{remark}
	The definition of the uniform observability of the pair $\left(A_{.},C_{.}\right)$ can be viewed as a milder condition to that of the time-invariant case. Firstly, consider the time-invariant system as a special case of the periodic system, i.e., $A_k\equiv A$ and $C_k\equiv C,\;\forall k\in\mathbb{N}$. Then the condition of observability of LTI system requires the matrix pair $(A_k,C_k)$ to be observable for each time step $k$. Consider the following periodic example with
	$$
	A_k\triangleq\begin{bmatrix}
	2&0\\0&2
	\end{bmatrix},\quad C_{2k-1}=[1,0],\;\;C_{2k}=[0,1],
	$$ 
	which represents a $2$-periodic linear system. At each time step $k$, the matrix pair $\left(A_{k},C_{k}\right)$ is not observable, while one can construct the feedback gain $L_{2k-1}=[-1.9,0]^T,L_{2k}=[0,-1.9]^T$ to make the monodromy matrix of the feedback system, i.e., $$\phi_k=\left(A_{k+1}+L_{k+1}C_{k+1}\right)\left(A_k+L_kC_k\right)=\begin{bmatrix}
	0.2&0\\0&0.2
	\end{bmatrix},$$ Schur-stable for any $k$. Thus, a uniformly observable matrix pair $\left(A_.,C_.\right)$ is not required to be observable for each $\left(A_k,C_k\right)$, which can be viewed as a milder condition to the time-invariant case.  
\end{remark}
}

\textcolor{black}{As proposed in \cite{battistelli2014kullback, kamal2013information, bittani1988}, the above two assumptions are actually mild for distributed filtering problems. Generally speaking, Assumption \ref{as1} is to ensure globally average weighting of the information from all the nodes of the sensor network so as to achieve the performance of  the centralized optimal case, and Assumption \ref{as2} is essential for the existence of the SPPS solution to the corresponding DPRE. Namely, Assumption \ref{as2} guarantees the stability of the centralized Kalman filtering algorithm on periodic systems \cite{bittani1988}.}
% It is worth mentioning that the doubly stochastic matrix $\mathcal{L}$ can also be obtained in a distributed way as long as the communication graph is directed and strongly connected \cite{GHARESIFARD2012539}.}
\subsection{Algorithm Convergence Analysis}
In this subsection, the performance of the proposed Algorithm \ref{alg1} on periodic systems will be presented.

\textcolor{black}{Generally speaking, the performance of periodic system can be evaluated via two main mathematical techniques, i.e., the time-invariant transformation in \cite{bittanti1990algebraic} and the DPRE in \cite{bittani1988}. Due to the complex coupling between the process noise and observation noise of the transformed system, and the sub-optimality of CMDF, the time-invariant transformation method needs to overcome the difficulties such as the reformulation of the relatively complex CMDF algorithm for periodic systems into LTI case, and the verification of the equivalence between solutions to different Riccati equations under sub-optimal case, which may be very complex. Hence, in this paper, the more intuitional DPRE is mainly used for performance analysis and some new properties of DPRE are excavated to facilitate the understanding of CMDF for periodic system.}

For the centralized Kalman filtering, Bittani {\it et al.} \cite{bittani1988} proved that, if the system $\left(A_.,C_.\right)$ is uniformly observable, then the iteration
$$
\begin{aligned}
P_{k+1|k}=&A_kP_{k|k-1}A_k^T+Q_k-A_kP_{k|k-1}C_k^T\\
&\times\big(C_kP_{k|k-1}C_k^T+R_k\big)^{-1}C_kP_{k|k-1}A_k^T
\end{aligned}
$$
converges to the SPPS solution to the DPRE
\begin{equation}\label{dare}
\begin{aligned}
P_{k+1}=&A_kP_kA_k^T+Q_k-A_kP_kC_k^T\\
&\times\big(C_kP_kC_k^T+R_k\big)^{-1}C_kP_kA_k^T
\end{aligned}
\end{equation}
with $k\to\infty$, where $P_{k+\mathcal{T}}=P_k$. Based on this premise, note that as the number of fusion step $L$ tends to infinity, each sensor node can precisely obtain the information $\sum_{i=1}^{N}C_{i,k}^TR_{i,k}^{-1}C_{i,k}$ and $\sum_{i=1}^{N}C_{i,k}^TR_{i,k}^{-1}y_{i,k}$ in a distributed way. Therefore, it is straightforward that the matrix $P_{i,k|k-1}$ converges to the centralized optimal SPPS performance with sufficient information fusion, i.e., 
$$
\lim_{k\to\infty}\lim_{L\to\infty}P_{i,k+1|k}-P_{k+1}=\textbf{O},\quad \forall i\in\mathcal{V}.
$$

\textcolor{black}{The objective of this subsection} is to first solve the essential problem induced by the time-variant property of the systems, then reveal the asymptotically periodic property of the parameter matrix $P_{i,k|k-1}$ and the estimation error covariance matrix of the CMDF algorithm on linear periodic system, with {a finite fusion step $L$.}

With Algorithm \ref{alg1}, the iteration of {$P_{i,k|k}^{-1}$} can be formulated as
\begin{equation}
\begin{aligned}
P_{i,k|k}^{-1}=P_{i,k|k-1}^{-1}+N\sum_{j=1}^{N}l_{ij}^{(L)}C_{j,k}^{T}R_{j,k}^{-1}C_{j,k}.
\end{aligned}
\end{equation}
{Let} the modified observation matrix for each sensor {be}
$$
\tilde{C}_{i,k}^{(L)}=\Big[\mathrm{sign}\big(l_{i1}^{(L)}\big) C_{1,k}^T,\cdots,\mathrm{sign}\big(l_{iN}^{(L)}\big)C_{N,k}^T\Big]^T,
$$
and the modified noise covariance matrix {be}
%$$
%\begin{aligned}
%    \big(\bar{R}_{i}^{(L)}\big)^{-1}&=diag\left(sign\big(l_{i1}^{(L)}\big)R_{1}^{-1},\cdots,sign\big(l_{iN}^{(L)}\big)R_{N}^{-1}\right)\\
%	\big(\tilde{R}_{i}^{(L)}\big)^{-1}&=diag\left(Nl_{i1}^{(L)}R_{1}^{-1},\cdots,Nl_{iN}^{(L)}R_{N}^{-1}\right),
%\end{aligned}
%$$
\textcolor{black}{
	$$
	\begin{aligned}
	\bar{R}_{i,k}^{(L)}&=\mathrm{diag}\left(\mathrm{sign}\big(l_{i1}^{(L)}\big)R_{1,k},\cdots,\mathrm{sign}\big(l_{iN}^{(L)}\big)R_{N,k}\right),\\
	\tilde{R}_{i,k}^{(L)}&=\mathrm{diag}\left(\frac{1}{Nl_{i1}^{(L)}}R_{1,k},\cdots,\frac{1}{Nl_{iN}^{(L)}}R_{N,k}\right),
	\end{aligned}
	$$
	where 
	$$
	\mathrm{sign}\left(x\right)=\left\{\begin{aligned}
	-&1,\quad x<0\\
	&0,\quad x=0\\
	&1,\quad x>0
	\end{aligned}\right.,
	$$ 
	and $\frac{1}{Nl_{i1}^{(L)}}$ is set to $0$ if the denominator $Nl_{ij}^{(L)}=0$.}
\textcolor{black}{Note that the term $\mathrm{sign}\big(l_{ij}^{(L)}\big)$ only takes the value of $0$ or $1$ due to the non-negativity of $\mathcal{L}^L$ and the function $\mathrm{sign}(\cdot)$ is essentially used as indicator function.} {The following Lemma describes} the asymptotically periodic property of the iteration $P_{i,k+1|k}$.
\vspace{5pt}
{\begin{lemma}\label{lmric}
		If $\big(A_.,\tilde{C}_{i,.}^{(L)}\big)$ is uniformly observable for all $i\in\mathcal{V}$, then $P_{i,k+1|k}$ converges to the SPPS solution of the DPRE 
		\begin{equation}\label{ricori}
		\begin{aligned}
		P_{i,k+1}^{(L)}=&A_kP_{i,k}^{(L)}A_k^{T}+Q_k-A_kP_{i,k}^{(L)}\big(\tilde{C}_{i,k}^{(L)}\big)^T\\
		&\times\Big(\tilde{C}_{i,k}^{(L)} P_{i,k}^{(L)}\big(\tilde{C}_{i,k}^{(L)}\big)^T+\tilde{R}_{i,k}^{(L)}\Big)^{-1}\tilde{C}_{i,k}^{(L)} P_{i,k}^{(L)}A_k^{T},
		\end{aligned}
		\end{equation}
		i.e.,
		$$
		\lim_{k\to \infty}\big(P_{i,k+1|k}-P_{i,k+1}^{(L)}\big)=\textbf{O},\qquad \forall i\in\mathcal{V}.
		$$
\end{lemma}}
\vspace{6pt}
\begin{proof}
	{With the definition of $\tilde{C}_{i,k}^{(L)}$ and $\tilde{R}_{i,k}^{(L)}$, the iteration of $P_{i,k+1|k}$ can be reformulated as}
	\begin{equation}\label{eqreformu}
	\begin{aligned}
	P_{i,k+1|k}=&A_k\left(P_{i,k|k-1}^{-1}+\big(\tilde{C}_{i,k}^{(L)}\big)^T\big(\tilde{R}_{i,k}^{(L)}\big)^{-1}\tilde{C}_{i,k}^{(L)}\right)^{-1}A_k^{T}\\ &+Q_k.
	\end{aligned}	
	\end{equation}
\vspace{6pt}	
	{With Lemma \ref{lm2}, the following equality is obtained:}
	{
		$$
		\begin{aligned}
		&\quad\Big(P_{i,k|k-1}^{-1}+\big(\tilde{C}_{i,k}^{(L)}\big)^T\big(\tilde{R}_{i,k}^{(L)}\big)^{-1}\tilde{C}_{i,k}^{(L)}\Big)^{-1}\\
		&=P_{i,k|k-1}-P_{i,k|k-1}\big(\tilde{C}_{i,k}^{(L)}\big)^T\\
		&\quad\times\Big(\tilde{C}_{i,k}^{(L)} P_{i,k|k-1}\big(\tilde{C}_{i,k}^{(L)}\big)^T+\tilde{R}_{i,k}^{(L)}\Big)^{-1}\tilde{C}_{i,k}^{(L)} P_{i,k|k-1}.
		\end{aligned}
		$$}
	{The equation \eqref{eqreformu} can be reformulated as
		$$
		\begin{aligned}
		&P_{i,k+1|k}=A_kP_{i,k|k-1}A_k^T+Q_k-A_kP_{i,k|k-1}\big(\tilde{C}_{i,k}^{(L)}\big)^T\\
		&\quad\times\Big(\tilde{C}_{i,k}^{(L)} P_{i,k|k-1}\big(\tilde{C}_{i,k}^{(L)}\big)^T+\tilde{R}_{i,k}^{(L)}\Big)^{-1}\tilde{C}_{i,k}^{(L)} P_{i,k|k-1}A_k^{T}.\\
		\end{aligned}
		$$}
	
	Thus, equation \eqref{ricori} is the SPPS form of equation \eqref{eqreformu}. With Theorem 7 in \cite{bittani1988}, if $\big(A_.,\tilde{C}_{i,.}^{(L)}\big)$ is uniformly observable, then $P_{i,k+1|k}$ converges to the solution of the DPRE \eqref{ricori}.
\end{proof}
\vspace{6pt}
% \begin{remark}
% 	If the number of fusion step is less than the diameter of the communication topology, i.e., $L<d$, \textcolor{blue}{then the matrix $\tilde{R}_i^{(L)}$ is not invertible since some $l_{ij}^{(L)}$ may be $0$. However, {one can eliminate the corresponding zero blocks} in $\tilde{C}_i^{(L)}$ and $\tilde{R}_i^{(L)}$ to make the modified $\tilde{R}_i^{(L)}$ invertible, or replace the inverse sign with generalized inverse, i.e., $\big(\tilde{R}_i^{(L)}\big)^{\dagger}$.  These two kinds of modification do not affect the observability of the pair $\big(A,\tilde{C}_i^{(L)}\big)$ and the equivalent relationship $$\tilde{C}_i^{(L)}\big(\tilde{R}_i^{(L)}\big)^{\dagger}\tilde{C}_i^{(L)}=N\sum_{j=1}^{N}l_{ij}^{(L)}C_{j}^{T}R_{j}^{-1}C_{j}.$$ 
%    {In order to simplify the notation, $\tilde{C}_i^{(L)}$, $\tilde{R}_i^{(L)}$, $\big(\tilde{R}_i^{(L)}\big)^{-1}$  and $\bar{R}_i^{(L)}$ will be kept to denote the modified observation and noise matrices.}}
% \end{remark}
% \vspace{6pt}

\textcolor{black}{It can be observed from DPRE \eqref{ricori} that the SPPS performance of the CMDF algorithm is closely related to the number of fusion step $L$ between two sampling instants, which is reflected in the expression of the modified matrices $\tilde{C}_{i,k}^{(L)}$ and $\tilde{R}_{i,k}^{(L)}$, and the performance degradation of CMDF compared with CKF is essentially induced by the mismatch of the matrix $\tilde{R}_{i,k}^{(L)}$.}
% reflecting on the value of the matrices $\mathcal{L}^{(L)}$ and $C_{i,k}^{(L)}$. 
\textcolor{black}{However, the matrix $P_{i,k|k}^{(L)}$ cannot represent the estimation error covariance matrix for each sensor $i$. Hence, in addition to the above analysis on $P_{i,k+1|k}$, {it also needs to analyze} the SPPS performance of the real estimation error covariance matrix.}

In order to formulate the asymptotically periodic property of error covariance matrix, the following Lemma is proposed for the description of the convergence property of Lyapunov iteration with asymptotically periodic parameters, \textcolor{black}{which is a theoretical improvement to Theorem 1 in \cite{cattivelli2010diffusion}.}

\vspace{6pt}
\begin{lemma}\label{lmlya}
	Consider the recursion of the form 
	$$
	P_{k+1}=A_kP_kA_k^T+Q_k
	$$
	where $A_k$ and $Q_k$ converge uniformly to $\bar{A}_k$ and $\bar{Q}_k$, respectively. $\bar{A}_k$ and $\bar{Q}_k$ are $\mathcal{T}$-periodic matrices, i.e., $\bar{A}_{k+\mathcal{T}}=\bar{A}_{k}$ and $\bar{Q}_{k+\mathcal{T}}=\bar{Q}_{k}$, $\forall k\in\mathbb{N}$. $\bar{A}_{.}$ is asymptotically stable, i.e., $\Pi_{i=1}^{\mathcal{T}}\bar{A}_i$ is Schur stable. Then, $P_{k+1}$ converges to the SPPS solution to the DPLE
	$$
	\bar{P}_{k+1}=\bar{A}_k\bar{P}_k\bar{A}_k^T+\bar{Q}_k,
	$$ i.e.
	$$
	\lim_{k\to\infty}\big(P_{k}-\bar{P}_k\big)=\textbf{O}.
	$$
\end{lemma}
\vspace{6pt}
The proof is given in Appendix \ref{prfLya}.
\vspace{6pt}

Denote the estimation error of sensor node $i$ as $\tilde{x}_{i,k|k-1}=x_k-\hat{x}_{i,k|k-1}$, $\tilde{x}_{i,k|k}=x_{k}-\hat{x}_{i,k|k}$, and the corresponding estimation error covariance matrix as $\tilde{P}_{i,k|k-1}=\mathbb{E}\big\lbrace\tilde{x}_{i,k|k-1}\tilde{x}_{i,k|k-1}^{T}\big\rbrace$, $\tilde{P}_{i,k|k}=\mathbb{E}\big\lbrace\tilde{x}_{i,k|k}\tilde{x}_{i,k|k}^{T}\big\rbrace$. The following theorem describes the asymptotically periodic property of $\tilde{P}_{i,k|k-1}$.
\vspace{6pt}
% $$
% \begin{aligned}
% 	&\tilde{x}_{i,k|k-1}=x_k-x_{i,k|k-1}\\
% 	&\tilde{x}_{i,k|k}=x_{k}-x_{i,k|k},
% \end{aligned}
% $$
% $$
% \begin{aligned}
% 	&\qquad\quad\tilde{P}_{i,k|k-1}=\mathbb{E}\left\lbrace\tilde{x}_{i,k|k-1}\tilde{x}_{i,k|k-1}^{T}\right\rbrace\\
% 	&\qquad\quad\tilde{P}_{i,k|k}=\mathbb{E}\left\lbrace\tilde{x}_{i,k|k}\tilde{x}_{i,k|k}^{T}\right\rbrace.
% \end{aligned}
% $$
{
	\begin{theorem}\label{thmconverge}
		If $\big(A_.,\tilde{C}_{i,.}^{(L)}\big)$ is uniformly observable for $i\in\mathcal{V}$, then the iteration of $\tilde{P}_{i,k|k-1}$ will converge to the SPPS solution to the discrete-time periodic Lyapunov equation (DPLE)
		\begin{equation}\label{lyaeq}
		\begin{aligned}
		&\tilde{P}_{i,k+1}^{(L)}=\tilde{A}_{P_{i,k}^{(L)}}\tilde{P}_{i,k}^{(L)}\tilde{A}_{P_{i,k}^{(L)}}^T+Q_k+K_{P_{i,k}^{(L)}}\bar{R}_{i,k}^{(L)}K_{P_{i,k}^{(L)}}^T,\\
		\end{aligned}
		\end{equation}
		where
		$$
		\begin{aligned}
		\tilde{A}_{P_{i,k}^{(L)}}\triangleq A_k-&A_kP_{i,k}^{(L)}\big(\tilde{C}_{i,k}^{(L)}\big)^{T}\\
		&\times\Big(\tilde{C}_{i,k}^{(L)}P_{i,k}^{(L)}\big(\tilde{C}_{i,k}^{(L)}\big)^T+\tilde{R}_{i,k}^{(L)}\Big)^{-1}\tilde{C}_{i,k}^{(L)},
		\end{aligned}
		$$ 
		$$
		\begin{aligned}
		\quad\; K_{P_{i,k}^{(L)}}\triangleq\; &A_kP_{i,k}^{(L)}\big(\tilde{C}_{i,k}^{(L)}\big)^T\\
		&\quad\;\times\Big(\tilde{C}_{i,k}^{(L)}P_{i,k}^{(L)}\big(\tilde{C}_{i,k}^{(L)}\big)^T+\tilde{R}_{i,k}^{(L)}\Big)^{-1}.\;\qquad\quad
		\end{aligned}
		$$
	\end{theorem}
	\vspace{6pt}
	\begin{proof}
	\textcolor{black}{With Algorithm \ref{alg1}, one obtains that
		$$
		\begin{aligned}
		\tilde{x}_{i,k|k}=&P_{i,k|k}P_{i,k|k}^{-1}x_k-P_{i,k|k}\\
		\times&\Big(P_{i,k|k-1}^{-1}\hat{x}_{i,k|k-1}+N\sum_{j=1}^{N}l_{ij}^{(L)}C_{j,k}^{T}R_{j,k}^{-1}y_{j,k}\Big)\\
		=&\;P_{i,k|k}\Big(P_{i,k|k-1}^{-1}\tilde{x}_{i,k|k-1}-\big(\tilde{C}_{i,k}^{(L)}\big)^T\big(\tilde{R}_{i,k}^{(L)}\big)^{-1}\tilde{v}_{i,k}\Big),\\
		\tilde{P}_{i,k|k}=&P_{i,k|k}P_{i,k|k-1}^{-1}\tilde{P}_{i,k|k-1}P_{i,k|k-1}^{-1}P_{i,k|k}\\
		+&P_{i,k|k}\big(\tilde{C}_{i,k}^{(L)}\big)^T\big(\tilde{R}_{i,k}^{(L)}\big)^{-1}\bar{R}_{i,k}^{(L)}\big(\tilde{R}_{i,k}^{(L)}\big)^{-1}\tilde{C}_{i,k}^{(L)}P_{i,k|k},
		\end{aligned}
		$$
		and
		$$
		\begin{aligned}
		\tilde{x}_{i,k+1|k}=&A_k\tilde{x}_{i,k|k}+\omega_{k},\\
		\tilde{P}_{i,k+1|k}=&A_k\tilde{P}_{i,k|k}A_k^T+Q_k,
		\end{aligned}
		$$
		where $\tilde{v}_{i,k}=\left[\mathrm{sign}(l_{i1}^{(L)})v_{1,k}^T,\cdots,\mathrm{sign}(l_{iN}^{(L)})v_{N,k}^T\right]^T$.
		{It follows that}
		$$
		\begin{aligned}
		A_kP_{i,k|k}P_{i,k|k-1}^{-1}&=A_k-A_kP_{i,k|k-1}\big(\tilde{C}_{i,k}^{(L)}\big)^{T}\\
		\times&\Big(\tilde{C}_{i,k}^{(L)}P_{i,k|k-1}\big(\tilde{C}_{i,k}^{(L)}\big)^T+\tilde{R}_{i,k}^{(L)}\Big)^{-1}\tilde{C}_{i,k}^{(L)},
		\end{aligned}
		$$
		and
		$$
		\begin{aligned}
		P_{i,k|k}\big(\tilde{C}_{i,k}^{(L)}\big)^{T}&\big(\tilde{R}_{i,k}^{(L)}\big)^{-1}=P_{i,k|k-1}\big(\tilde{C}_{i,k}^{(L)}\big)^{T}\\
		&\quad\;\;\times\Big(\tilde{C}_{i,k}^{(L)}P_{i,k|k-1}\big(\tilde{C}_{i,k}^{(L)}\big)^{T}+\tilde{R}_{i,k}^{(L)}\Big)^{-1}.\qquad\quad
		\end{aligned}
		$$
		As $k$ tends to infinity, these two matrices will converge to the periodic form, i.e.,
		$$
		\begin{aligned}
		\lim_{k\to\infty}&A_kP_{i,k|k}P_{i,k|k-1}^{-1}=\tilde{A}_{P_{i,k}^{(L)}}\\
		\lim_{k\to\infty}&A_kP_{i,k|k}\tilde{C}_{i,k}^T\big(\tilde{R}_{i,k}^{(L)}\big)^{-1}=K_{P_{i,k}^{(L)}}.
		\end{aligned}
		$$  
		{With $P_{i,k}^{(L)}$ being the solution to DPRE, the above matrix $\tilde{A}_{P_{i,k}^{(L)}}$ is asymptotically stable}, i.e., the spectral radius of $\prod_{j=1}^{\mathcal{T}}\tilde{A}_{P_{i,k+j}^{(L)}}$ is less than $1$ for all $k$ \cite{bittani1988}.
		Thus, with Lemma \ref{lmlya}, the iteration of $\tilde{P}_{i,k|k-1}$ will also asymptotically converge to the SPPS form, i.e., $\lim_{k\to\infty}\tilde{P}_{i,k|k-1}-\tilde{P}_{i,k}^{(L)}=\textbf{O}$, and the convergent solution  satisfies the DPLE \eqref{lyaeq}.}
	\end{proof}}
\vspace{6pt}
\begin{remark}\label{rmkThmConvergence}
		Theorem \ref{thmconverge} demonstrates that for CMDF, the performance degradation compared with the centralized optimal case is induced by {the imprecise values of 
			$\sum_{i=1}^{N} C_{i,k}^{T}R_{i,k}^{-1}y_{i,k}$ and $\sum_{i=1}^{N}C_{i,k}^TR_{i,k}^{-1}C_{i,k}$} in each sampling instant. 
		For periodic system, the effect of imprecise information on the state estimation performance will accumulate, and finally reflect on the SPPS solution to the corresponding DPRE (\ref{ricori}) and DPLE (\ref{lyaeq}), respectively. \textcolor{black}{Theorem 1 also indicates the least requirement for the stability of the CM-based distributed filtering algorithm on periodic systems, i.e., the uniform observability for the matrix pair $(A_.,\tilde{C}_{i,.}^{(L)})$ for each $i\in\mathcal{V}$, which also indicates that the fusion step can be smaller than the diameter $d$ of the graph $\mathcal{G}$ if the observed information from the $L$-step neighbors for each sensor $i$ is sufficient to guarantee the asymptotically periodic of the matrix iterative law in Algorithm \ref{alg1}. Such a condition can be checked through many kinds of criteria proposed in \cite{bittanti1985discrete}.}
\end{remark}
\vspace{6pt}

\section{Performance Gap Analysis}\label{bridge}
In this section, the SPPS performance gap between the CMDF and the CKF will be analyzed.

Note that the fusion step $L$ significantly influence the observability of $(A_.,\tilde{C}_{i,.}^{(L)})$, and the gap between $R_{k}$ and $\tilde{R}_{i,k}^{(L)}$, for any $i\in\mathcal{V}$. If the fusion step is larger than the diameter of the communication graph, i.e., $L\ge d$, then one has $l_{ij}^{(L)}>0$ for all $i,j$ \cite{battistelli2014kullback} and $\tilde{C}_{i,k}^{(L)}=C_{k}$ for each $i\in\mathcal{V}$. {Thus, one can} rewrite the DPRE (\ref{ricori}) for each sensor node as
\begin{equation}\label{redare}
\begin{aligned}
P_{i,k+1}^{(L)}=&A_kP_{i,k}^{(L)}A_{k}^{T}+Q_k\\
-&A_kP_{i,k}^{(L)}C_k^{T}\big(C_k P_{i,k}^{(L)}C_k^T+\tilde{R}_{i,k}^{(L)}\big)^{-1}C_k P_{i,k}^{(L)}A_k^{T},\\
\end{aligned}
\end{equation}
\textcolor{black}{where the performance degradation for CMDF is further reformulated as the mismatch of noise covariance matrix $R_k$ to $\tilde{R}_{i,k}^{(L)}$ for each sensor $i$.} In this subsection, the effect of such a mismatched $R_.$ on the properties of the SPPS solutions $P_{i,k}^{(L)}$ and $\tilde{P}_{i,k}^{(L)}$ with {$L\ge d$  will be discussed.}

Before discussing the property of the filtering performance gap, the following {two lemmas are established} to reveal the properties of the SPPS solutions to DPRE, which are of vital importance for the derivation of the main theorems. 
{For simplicity}, use the term $\mathrm{dpre}\left(A_.,C_.,Q_.,R_.\right)$ to denote the solution $P_.$ to DPRE (\ref{dare}). 
\vspace{6pt}
\begin{lemma}\label{lmmono}
	{The SPPS solution $P_.$ to the DPRE \eqref{dare} is monotonically increasing with $R_.$}. In other words, if $R_{1,k}\ge R_{2,k}>0,\,\forall k\in\mathbb{N}$, $P_{1,k}=\mathrm{dpre}\left(A_.,C_.,Q_.,R_{1,.}\right)$ and $P_{2,k}=\mathrm{dpre}\left(A_.,C_.,Q_.,R_{2,.}\right)$, then $P_{1,k}\ge P_{2,k},\;\forall k\in\mathbb{N}$.
\end{lemma}
\vspace{6pt}
The proof is given in Appendix \ref{pfmono}.

\vspace{6pt}
{The feedback system matrix at each time step $k$ is denoted as $\tilde{A}_{P_k}=A_k-A_kP_kC_k^T$ $\left(C_kP_kC_k^T+R_k\right)^{-1} C_k$, which may not be Schur-stable for periodic system. However, the {\it monodromy matrix}, denoted as $\Phi_{P_k}=\prod_{i=1}^{\mathcal{T}}\tilde{A}_{P_{k+i}}$, is Schur stable if $\big(A_.,C_.\big)$ is uniformly observable \cite{bittani1988}. Based on these results, the following lemma describes the properties of the eigenvalue and singularvalue of $\Phi_{P_k}$.
	\vspace{5pt}
	{\begin{lemma}\label{lmstable}
			The matrix $\Phi_{P_k}$ corresponding to the solution of DPRE \eqref{dare} is Schur stable and 
			$$
			\begin{aligned}
			\rho\big(\Phi_{P_k}\big)\leq \sqrt{1-\frac{\lambda_{min}\left(Q_.\right)}{\lambda_{max}\left(P_.\right)}}\;,\quad
			\big\|\Phi_{P_k}\big\|_2\leq \sqrt{\frac{\lambda_{max}\left(P_.\right)}{\lambda_{min}\left(Q_.\right)}},
			\end{aligned}
			$$ 
			where $\lambda_{min}(Q_.)$ is the smallest eigenvalue of $Q_k$ and $\lambda_{max}(P_.)$ is the largest eigenvalue of $P_k$, for all $k\in\mathbb{N}$.
		\end{lemma}
		\vspace{6pt}
		The proof is given in Appendix \ref{Pflmstable}.
	}
	\vspace{6pt}
	\begin{remark}
		It is illustrated in Lemma \ref{lmstable} that the spectral radius and 2-norm of $\Phi_{P_k}$ are bounded by the eigenvalues of SPPS solution $P_k$ to the DPRE \eqref{dare} and the noise matrix $Q_k$. This property is of vital importance to analyze the uniform property of the solutions to a group of DPRE, {as seen} in the proof of Theorem \ref{thmric}.
	\end{remark}
	\vspace{6pt}

	\vspace{6pt}
	{\begin{theorem}\label{thmric}
			For any given $\mathcal{T}$-periodic system matrices $\left(A_.,C_.\right)$ and $Q_.,R_.$, {there exist two constants} $M_1>0$ and $0<q_1<1$ such that
			$$
			\big\|P_k-P_{i,k}^{\left(L\right)}\big\|_2\leq M_1q_1^L, \quad\forall i\in\mathcal{V},\, L\ge d,\, k\in\mathbb{N},
			$$
			where $P_k$ and $P_{i,k}^{(L)}$ are the SPPS solutions to DPRE \eqref{dare} and \eqref{redare}, respectively.
		\end{theorem}
		\vspace{6pt}
		\begin{proof}
	    \textcolor{black}{Denote the SPPS performance of the $i$-th sensor and centralized optimal case as $P_{i,k}^{(L)}$, $P_{k}$, respectively, where $P_{i,k+\mathcal{T}}^{(L)}=P_{i,k}^{(L)}$ and $P_{k+\mathcal{T}}=P_{k}$.}
	    
	    \textcolor{black}{{First,} by some calculations, one can reformulate the iteration form of the performance gap as
	    	\begin{equation}\label{pfgap}
	    	\begin{aligned}
	    	P_{i,k+1}^{(L)}-P_{k+1}&=\tilde{A}_{P_{i,k}^{(L)}}\big(P_{i,k}^{(L)}-P_{k}\big)\tilde{A}_{P_{k}}^T\\
	    	&\qquad\qquad+K_{P_{i,k}^{(L)}}\big(\tilde{R}_{i,k}^{(L)}-R_k\big)K_{P_k}^T,
	    	\end{aligned}
	    	\end{equation}  
	    	where
	    	$$
	    	\begin{aligned}
	    	K_{P_{i,k}^{(L)}}&=A_kP_{i,k}^{(L)}C_{k}^T\big(C_kP_{i,k}^{(L)}C_k^T+R_{i,k}^{(L)}\big)^{-1},\\
	    	K_{P_{k}}&=A_kP_{k}C_{k}^T\big(C_kP_{k}C_k^T+R_{k}\big)^{-1},\\
	    	\tilde{A}_{P_{i,k}^{(L)}}&=A_k-K_{P_{i,k}^{(L)}}C_k,\\
	    	\tilde{A}_{P_{k}}&=A_k-K_{P_k}C_k.
	    	\end{aligned}
	    	$$
	    	Perform the iteration \eqref{pfgap} for $\mathcal{T}$-times, one has
	    	\begin{equation}\label{Tgap}
	    	\begin{aligned}
	    	P_{i,k+\mathcal{T}}^{(L)}-P_{k+\mathcal{T}}=\Phi_{P_{i,k}^{(L)}}\big(P_{i,k}^{(L)}-P_{k}\big)\Phi_{P_k}^T
	    	+\Psi_{P_{i,k}^{(L)}},
	    	\end{aligned}
	    	\end{equation}
	    	where
	    	$$
	    	\begin{aligned}
	    	\Phi_{P_{i,k}^{(L)}}&=\prod_{l=0}^{\mathcal{T}-1}\tilde{A}_{P_{i,k+l}^{(L)}},\quad \tilde{\Phi}_{i,k+\mathcal{T},k+l+1}^{(L)}=\prod_{p=l+1}^{\mathcal{T}-1}\tilde{A}_{P_{i,k+p}^{(L)}},\\
	    	\Phi_{P_{k}}&=\prod_{l=0}^{\mathcal{T}-1}\tilde{A}_{P_{k}},\quad \tilde{\Phi}_{k+\mathcal{T},k+l+1}=\prod_{p=l+1}^{\mathcal{T}-1}\tilde{A}_{P_{k+p}},\\
	    	\Psi_{P_{i,k}^{(L)}}&=\sum_{l=0}^{\mathcal{T}-1} \tilde{\Phi}_{i,k+\mathcal{T},k+l+1}^{(L)}
	    	K_{P_{i,k+l}^{(L)}}\\
	    	&\qquad\times\big(\tilde{R}_{i,k+l}^{(L)}-R_{k+l}\big)K_{P_{k+l}}^T\tilde{\Phi}_{k+\mathcal{T},k+l+1}^T.
	    	\end{aligned}
	    	$$
	        Using the Kalman equality that
	    	$$
	    	\begin{aligned}
	    	K_{P_{k}}&=A_k\bar{P}_kC_k^TR_k^{-1},\;K_{P_{i,k}^{(L)}}=A_k\bar{P}_{i,k}^{(L)}C_k^T\big(\tilde{R}_{i,k}^{(L)}\big)^{-1},\\
	    	\bar{P}_k&\triangleq\big(P_{k}^{-1}+C_k^TR_k^{-1}C_k\big)^{-1},\\
	    	\bar{P}_{i,k}^{(L)}&\triangleq\big(\big(P_{i,k}^{(L)}\big)^{-1}+C_k^T\big(\tilde{R}_{i,k}^{(L)}\big)^{-1}C_k\big)^{-1},
	    	\end{aligned}
	    	$$
	    	one can rewrite the expression of $\Psi_{P_{i,k}^{(L)}}$ as
	    	$$
	    	\begin{aligned}
	    	&\Psi_{P_{i,k}^{(L)}}=\sum_{l=0}^{\mathcal{T}-1} \tilde{\Phi}_{i,k+\mathcal{T},k+l+1}^{(L)}
	    	A_{k+l}\bar{P}_{i,k+l}^{(L)}C_{k+l}^T\\
	    	&\quad\times\big(R_{k+l}^{-1}-\big(\tilde{R}_{i,k+l}^{(L)}\big)^{-1}\big)C_{k+l}\bar{P}_{k+l}A_{k+l}^T\tilde{\Phi}_{k+\mathcal{T},k+l+1}^T.
	    	\end{aligned}
	    	$$}
	    
	    \textcolor{black}{Performing the iteration \eqref{Tgap} for $h$-times, together with the periodicity $P_{i,k+\mathcal{T}}^{(L)}=P_{i,k}^{(L)}$, $P_{k+\mathcal{T}}=P_k$, $\Phi_{P_{i,k+\mathcal{T}}^{(L)}}=\Phi_{P_{i,k}^{(L)}}$ and  $\Psi_{P_{i,k+\mathcal{T}}^{(L)}}=\Psi_{P_{i,k}^{(L)}}$, one has
	    	\begin{equation}\label{Seriegap}
	    	\begin{aligned}
	    	P_{i,k+h\mathcal{T}}^{(L)}-P_{k+h\mathcal{T}}&=\Phi_{P_{i,k}^{(L)}}^h\big(P_{i,k}^{(L)}-P_{k}\big)\big(\Phi_{P_k}^h\big)^T\\
	    	&\qquad\qquad+\sum_{j=0}^{h-1}\Phi_{P_{i,k}^{(L)}}^{j}\Psi_{P_{i,k}^{(L)}}\big(\Phi_{P_k}^j\big)^T.
	    	\end{aligned}
	    	\end{equation}
	    	Note that $\Phi_{P_{i,k}^{(L)}}$ and $\Phi_{P_k}$ are Schur stable, one has
	    	$$
	    	P_{i,k}^{(L)}-P_{k}=\sum_{j=0}^{\infty}\Phi_{P_{i,k}^{(L)}}^{j}\Psi_{P_{i,k}^{(L)}}\big(\Phi_{P_k}^j\big)^T,
	    	$$
	    	and 
	    	$$
	    	\begin{aligned}
	    	\big\|P_{i,k}^{(L)}-P_{k}\big\|_2&\leq\big\|\Psi_{P_{i,k}^{(L)}}\big\|_2\sum_{j=0}^{\infty}\big\|\Phi_{P_{i,k}^{(L)}}^{j}\big\|_2\big\|\big(\Phi_{P_k}^j\big)^T\big\|_2\\
	    	&=\big\|\Psi_{P_{i,k}^{(L)}}\big\|_2\sum_{j=0}^{\infty}\big\|\Phi_{P_{i,k}^{(L)}}^{j}\big\|_2\big\|\Phi_{P_k}^j\big\|_2,
	    	\end{aligned}
	    	$$
	    	which are the infinite series form of the performance gap $P_{i,k}^{(L)}-P_{k}$ and $\big\|P_{i,k}^{(L)}-P_{k}\big\|_2$, respectively.}
	    
	    \textcolor{black}{Consider the term $R_{k+l}^{-1}-\big(\tilde{R}_{i,k+l}^{(L)}\big)^{-1}$ in $\Psi_{P_{i,k}^{(L)}}$. By Lemma \ref{lm1}, the matrix $\mathcal{L}^{L}$ converges to $\frac{1}{N}\textbf{1}_N\textbf{1}_N^T$ exponentially {with increasing $L$}. Hence, one can find two parameters $0<q_1<1$ and $M_1^{' }>0$ such that
	    	$$
	    	\big\|R_{k}^{-1}-\big(\tilde{R}_{i,k}^{(L)}\big)^{-1}\big\|_2<M_1^{'}q_1^L,\;\forall i\in\mathcal{V},\;L\ge d,\;k\in\mathbb{N}.
	    	$$
	    	Note that for the term $\Psi_{P_{i,k}^{(L)}}$, one has
	    	$$
	    	\begin{aligned}
	    	\big\|\Psi_{P_{i,k}^{(L)}}\big\|_2\leq& M_1^{''}q_1^L\sum_{l=0}^{\mathcal{T}-1} \big\|\tilde{\Phi}_{i,k+\mathcal{T},k+l+1}^{(L)}\big\|_2\big\|
	    	\tilde{\Phi}_{k+\mathcal{T},k+l+1}\big\|_2\\
	    	&\times\big\|\bar{P}_{i,k+l}^{(L)}\big\|_2\big\|\bar{P}_{k+l}\big\|_2,
	    	\end{aligned}
	    	$$ 
	    	where the inequality holds due to the uniform boundedness of $A_.$ and $C_.$.
	    	For the state transition matrix $\tilde{\Phi}_{i,k+\mathcal{T},k+l+1}^{(L)}$, using the same technique in the proof of Lemma \ref{lmstable}, one can verify that the following inequality:
	    	$$
	    	\big\|\tilde{\Phi}_{i,k+\mathcal{T},k+l+1}^{(L)}\big\|_2\leq\sqrt{\frac{\lambda_{max}\big(P_{i,.}^{(L)}\big)}{\lambda_{min}\big(Q_.\big)}},\quad\forall k\in\mathbb{N},\;l<\mathcal{T},
	    	$$ 
	    	also holds. Choose a sufficiently large $\mathcal{T}$-periodic positive-definite matrix sequence $\mathcal{R}_.$ such that $R_{i,k}^{(L)}\leq\mathcal{R}_k$ for all $i\in\mathcal{V}$, $L\ge d$ and $k\in\mathbb{N}$. Denote $\mathcal{P_.}=\mathrm{dpre}\left(A_.,C_.,Q_.,\mathcal{R}_.\right)$. 
	    	In virtue of Lemma \ref{lmmono} and Lemma \ref{lmstable}, one can obtain that 
	    	$$
	    	\begin{aligned}
	    	\big\|\tilde{\Phi}_{i,k+\mathcal{T},k+l+1}^{(L)}\big\|_2&\leq\sqrt{\frac{\lambda_{max}\big(\mathcal{P}_.\big)}{\lambda_{min}\big(Q_.\big)}},\\
	    	\big\|\bar{P}_{i,k+l}^{(L)}\big\|_2&\leq \big\|P_{i,k+l}^{(L)}\big\|_2\leq \lambda_{max}\big(\mathcal{P}_.\big).
	    	\end{aligned}
	    	$$ 
	    	Thus one can obtain a uniform bound for $\big\|\Psi_{P_{i,k}^{(L)}}\big\|_2$, i.e.,
	    	$$
	    	\big\|\Psi_{P_{i,k}^{(L)}}\big\|_2\leq Mq_1^L,
	    	$$
	    	where $M$ is not related to the sensor number $i$ and the fusion step $L$.}
	    
	    \textcolor{black}{In addition to the above analysis, the uniform boundedness of $\sum_{j=0}^{\infty}\big\|\Phi_{P_{i,k}^{(L)}}^{j}\big\|_2\big\|\Phi_{P_k}^j\big\|_2$ also needs to be proved to guarantee the exponential convergence of $\big\|P_{i,k}^{(L)}-P_k\big\|_2$ with the increase of $L$.}
	    
	    \textcolor{black}{Using Lemma \ref{lm3}, one can obtain that
	    	\begin{small}
	    		$$
	    		\Big\|\Phi_{P_{i,k}^{(L)}}^l\Big\|_2\leq\sqrt{n}\sum_{j=0}^{n-1}\binom{n-1}{j}\binom{l}{j}\Big\|\Phi_{P_{i,k}^{(L)}}\Big\|_2^j\rho\Big(\Phi_{P_{i,k}^{(L)}}\Big)^{l-j},
	    		$$
	    \end{small}
         \textcolor{black}{where} $n$ is the dimension of the system matrix $A_.$. With the pre-obtained matrix sequence $\mathcal{P}_.$,}
	    \textcolor{black}{one can obtain that 
	    	$$
	    	\begin{aligned}
	    	&\Big\|\Phi_{P_{i,k}^{(L)}}\Big\|_2\leq \sqrt{\frac{\lambda_{max}\big(P_{i,.}^{(L)}\big)}{\lambda_{min}\left(Q_.\right)}}\leq\sqrt{\frac{\lambda_{max}\left(\mathcal{P}_.\right)}{\lambda_{min}\left(Q_.\right)}}\triangleq\sigma_{1\mathcal{P}},\\
	    	&\rho\big(\Phi_{P_{i,k}^{(L)}}\big)\leq \sqrt{1-\frac{\lambda_{min}\left(Q_.\right)}{\lambda_{max}\big(P_{i,.}^{(L)}\big)}}\leq\sqrt{1-\frac{\lambda_{min}\left(Q_.\right)}{\lambda_{max}\left(\mathcal{P}_.\right)}}\triangleq \rho_{\mathcal{P}}.
	    	\end{aligned}
	    	$$
	    	Thus, one has
	    	$$
	    	\begin{aligned}
	    	\sum_{l=0}^{\infty}\Big\|\Phi_{P_{i,k}^{(L)}}^l\Big\|_2\leq&\sum_{l=0}^{\infty}\sqrt{n}\sum_{j=0}^{n-1}\binom{n-1}{j}\binom{l}{j}\sigma_{1\mathcal{P}}^j\rho_{\mathcal{P}}^{l-j}\\
	    	\leq&\sum_{l=0}^{\infty}\sqrt{n}\left(n\max_{j}\binom{n-1}{j}\sigma_{1\mathcal{P}}^j\right)l^n\rho_{\mathcal{P}}^{l-n}
	    	\end{aligned}
	    	$$
	    	for all $i\in\mathcal{V}$, $L\ge d$ and $k\in\mathbb{N}$. It is {easy} to verify the convergence of the infinite sum $\sum_{l=0}^{\infty}\big\|\Phi_{P_{i,k}^{(L)}}^l\big\|_2$ with $\rho_{\mathcal{P}}<1$. {Based on} the fact that the 2-norm $\big\|\Phi_{P_{k}}^l\big\|_2$ is uniformly bounded for all $l$, one can find a real number $M_0^{'}$ such that
	    	$$
	    	\sum_{l=0}^{\infty}\big\|\Phi_{P_{i,k}^{(L)}}^{l}\big\|_2\big\|\Phi_{P_k}^l\big\|_2\leq M_0^{'},\;\;\forall i\in\mathcal{V},\;L\ge d,\;k\in\mathbb{N}.
	    	$$
	    	With the above analysis, one only need to let $M_1=MM_0^{'}$ and the proof is completed.}
        \end{proof}
	\vspace{6pt}
	\begin{remark}
		The main idea of the proof of Theorem \ref{thmric} is to formulate the infinite series form of the performance gap $P_k-P_{i,k}^{(L)}$ based on the structure of DPRE. 
		The main difficulties of proving Theorem \ref{thmric} lie in two parts: one is to formulate a uniform upper bound of $\sum_{j=0}^{\infty}\big\|\Phi_{P_k}^j\big\|_2\big\|\Phi_{P_{i,k}^{(L)}}^j\big\|_2$ for all $i\in\mathcal{V},\;L\ge d,\;k\in\mathbb{N}$, and the other is to depict the uniformly linear relation between $\big\|\Psi_{P_{i,k}^{(L)}}\big\|_2$ and $\big\|R_{k}^{-1}-\big(\tilde{R}_{i,k}^{(L)}\big)^{-1}\big\|_2$. Compared with the time-invariant case in \cite{qian2021consensusbased}, the time-varying parameters of the target system, the nonlinearity and the time-variant property of DPRE significantly  
		%	existence of the term $\tilde{\Phi}_{P_{i,k+\mathcal{T},k+l+1}^{(L)}}$ in $\Psi_{P_{i,k}^{(L)}}$
		increase the difficulty of the proof.
		%	Hence, the technique proposed in Lemma \ref{lmstable} can be used to address the aforementioned issues by bounding the 2-norm of the state transition matrix $\tilde{\Phi}_{P_{i,k+\mathcal{T},k+l+1}^{(L)}}$ with the solution to the DPRE.
		Hence, the theoretical results in Lemma \ref{lmstable} and Theorem \ref{thmric} not only improve the corresponding theory of LTI case \cite{qian2021consensusbased} to periodic systems, but also supplement the technique in the analysis of periodic systems.
	\end{remark}
	\vspace{6pt}
	
	It is apparent that the matrix $P_{i,k}^{(L)}$ cannot depict the SPPS performance of the real error covariance matrix, i.e., 
	$$
	\lim_{k\to\infty}\big(P_{i,k}^{(L)}-\mathbb{E}\big\lbrace\tilde{x}_{i,k|k-1}\tilde{x}_{i,k|k-1}^T\big\rbrace\big)\neq\textbf{O}.
	$$ 
	With the result of Theorem \ref{thmconverge}, the iteration of the real covariance $\mathbb{E} \big\lbrace\tilde{x}_{i,k|k-1} \tilde{x}_{i,k|k-1}^T\big\rbrace$ will finally converge to the SPPS solution to the DPLE \eqref{lyaeq}. Thus the performance degradation of $\mathbb{E}\big\lbrace\tilde{x}_{i,k|k-1}\tilde{x}_{i,k|k-1}^T\big\rbrace$ needs further discussion on the effect of insufficient information fusion.
	
	{First, the aim is to} analyze the gap between the matrix $P_{i,k}^{(L)}$ and the matrix $\tilde{P}_{i,k}^{(L)}$. 
	\vspace{6pt}
	{\begin{theorem}\label{thmcov}
			For any given $\mathcal{T}$-periodic system matrices $\left(A_.,C_.\right)$ and $Q_.,R_.$, there exist two constants $M_2>0$ and $0<q_2<1$ such that
			$$
			\big\|\tilde{P}_{i,k}^{(L)}-P_{i,k}^{\left(L\right)}\big\|_2\leq M_2q_2^L, \quad\forall i\in\mathcal{V},\;L\ge d,\;k\in\mathbb{N}.
			$$
		\end{theorem}
		\vspace{6pt}
		\begin{proof}
			\textcolor{black}{Note that the periodic matrix sequences $P_{i,k}^{(L)}$ and $\tilde{P}_{i,k}^{(L)}$ satisfy the following equations:
				$$
				\begin{aligned}
				&P_{i,k+1}^{(L)}=\tilde{A}_{P_{i,k}^{(L)}}P_{i,k}^{(L)}\tilde{A}_{P_{i,k}^{(L)}}^T+Q_k+K_{P_{i,k}^{(L)}}\tilde{R}_{i,k}^{(L)}K_{P_{i,k}^{{(L)}}}^T,\\
				&\tilde{P}_{i,k+1}^{(L)}=\tilde{A}_{P_{i,k}^{(L)}}\tilde{P}_{i,k}^{(L)}\tilde{A}_{P_{i,k}^{(L)}}^T+Q_k+K_{P_{i,k}^{(L)}}R_kK_{P_{i,k}^{{(L)}}}^T.
				\end{aligned}
				$$
				Then, one can obtain that
				\begin{equation}\label{Lyagap}
				\begin{aligned}
				&\tilde{P}_{i,k+\mathcal{T}}^{(L)}-P_{i,k+\mathcal{T}}^{(L)}=\Phi_{P_{i,k}^{(L)}}\big(\tilde{P}_{i,k}^{(L)}-P_{i,k}^{(L)}\big)\Phi_{P_{i,k}^{(L)}}^T
				+\Upsilon_{P_{i,k}^{(L)}},\\
				&\Upsilon_{P_{i,k}^{(L)}}=\sum_{l=0}^{\mathcal{T}-1} \tilde{\Phi}_{i,k+\mathcal{T},k+l+1}^{(L)}
				K_{P_{i,k+l}^{(L)}}\\
				&\qquad\qquad\times\big(R_{k+l}-\tilde{R}_{i,k+l}^{(L)}\big)K_{P_{i,k+l}^{(L)}}^T\big(\tilde{\Phi}_{i,k+\mathcal{T},k+l+1}^{(L)}\big)^T.
				\end{aligned}
				\end{equation}
				where $\Phi_{P_{i,k}^{(L)}}$ and $\tilde{\Phi}_{i,k+\mathcal{T},k+l+1}^{(L)}$ are defined in the proof of Theorem \ref{thmric}.}
			
			\textcolor{black}{With $K_{P_{i,k}^{(L)}}=A_k\bar{P}_{i,k}^{(L)}C_k^T\big(\tilde{R}_{i,k}^{(L)}\big)^{-1}$ and similarly to the proof of Theorem \ref{thmric}, one can obtain an upper bound of the 2-norm of $\Upsilon_{P_{i,k}^{(L)}}$ as
				$$
				\begin{aligned}
				\big\|\Upsilon_{P_{i,k}^{(L)}}\big\|_2\leq& M_2^{'}\sum_{l=0}^{\mathcal{T}-1} \big\|\big(\tilde{R}_{i,k+l}^{(L)}\big)^{-1}\\
				&\qquad\times\big(R_{k+l}-\tilde{R}_{i,k+l}^{(L)}\big)^{-1}\big(\tilde{R}_{i,k+l}^{(L)}\big)^{-1}\big\|_2^2.
				\end{aligned}
				$$ }
			
			\textcolor{black}{Both matrices $R_k$ and $\tilde{R}_{i,k+l}^{(L)}$ are block diagonal. Thus, one can obtain that
				$$
				\begin{aligned}
				\Lambda^{(L)}R_{k}^{-1}=\big(\tilde{R}_{i,k}^{(L)}\big)^{-1}\big(R_{k}-\tilde{R}_{i,k}^{(L)}\big)^{-1}\big(\tilde{R}_{i,k}^{(L)}\big)^{-1},\\
				\end{aligned}
				$$
				where $\Lambda^{(L)}=\mathrm{diag}\big(\tilde{l}_{i1}^{(L)}I_{n_1},\cdots,\tilde{l}_{iN}^{(L)}I_{n_N}\big)$ and $\tilde{l}_{ij}^{(L)}=N^2\big(l_{ij}^{(L)}\big)^2-Nl_{ij}^{(L)}$.}
			
			\textcolor{black}{Rewrite the expression of $\tilde{l}_{ij}^{(L)}$ as 
				$$
				\tilde{l}_{ij}^{(L)}=\big(Nl_{ij}^{(L)}-1\big)^2+\big(Nl_{ij}^{(L)}-1\big),
				$$
				and using Lemma \ref{lm1}, the term $Nl_{ij}^{(L)}-1$ converges to $0$ exponentially with increasing $L$, i.e., there exists $M_3>0$, $0<q_3<1$, such that $\forall i,j$, one has $\big|Nl_{ij}^{(L)}-1\big|\leq M_3q_3^L$. {For $L\ge log_{\frac{1}{q_3}}M_3$, one has} $\big(Nl_{ij}^{(L)}-1\big)^2<\big|Nl_{ij}^{(L)}-1\big|$, {therefore there exist two parameters} $M_2^{''}\ge 0$ and $0<q_2<1$ such that
				$$
				\begin{aligned}
				\Big\|\big(\tilde{R}_{i,k}^{(L)}\big)^{-1}\big(R_k-\tilde{R}_{i,k}^{(L)}\big)\big(\tilde{R}_{i,k}^{(L)}\big)^{-1}\big\|_2\leq M_2^{''}q_2^{L}.
				\end{aligned}
				$$
				Hence, the exponential bound of $\big\|\Upsilon_{P_{i,k}^{(L)}}\big\|_2$ is accordingly obtained as
				$$
				\big\|\Upsilon_{P_{i,k}^{(L)}}\big\|_2\leq M_2^{'}q_2^L,\;\forall i\in\mathcal{V},\;L\ge d,\; k\in\mathbb{N}.
				$$}
			
			\textcolor{black}{Note that the infinite series form of the performance gap $\tilde{P}_{i,k}^{(L)}-P_{i,k}^{(L)}$ can be formulated as
				\begin{equation}\label{sumCov}
				\begin{aligned}
				\tilde{P}_{i,k}^{(L)}-P_{i,k}^{(L)}=\sum_{j=0}^{\infty}\Phi_{P_{i,k}^{(L)}}^{j}\Upsilon_{P_{i,k}^{(L)}}\big(\Phi_{P_{i,k}^{(L)}}^{j}\big)^T,
				\end{aligned}
				\end{equation}}
			
			\textcolor{black}{Similar to the proof of Theorem \ref{thmric}, one can obtain an uniform upper bound of $\sum_{j=0}^{\infty}\big\|\Phi_{P_{i,k}^{(L)}}^{j}\big\|_2^2\;$ for all $i\in\mathcal{V},\;L\ge d,\;k\in\mathbb{N}\;$.
				With the above analysis, the theorem is finally proved.}
		\end{proof}
	}
	\vspace{6pt}
	
	\textcolor{black}{The following corollary finally depicts the convergence property of the performance gap of $\tilde{P}_{i,k}^{(L)}$ compared with the centralized SPPS performance $P_k$, and the relationship between the decay rate $q$ and the spectral property of the gain matrix $\mathcal{L}$.}
	\vspace{6pt}
	\begin{corollary}\label{exponential}
		For any given $\mathcal{T}$-periodic system matrices $\left(A_.,C_.\right)$ and $Q_.,R_.$, there exist two constants $M_3>0$ and $0<q_3<1$ such that
		$$
		\big\|\tilde{P}_{i,k}^{(L)}-P_k\big\|_2\leq M_3q_3^L, \quad\forall i\in\mathcal{V},\;L\ge d,\;k\in\mathbb{N}.
		$$
		In addition, the paramter $q_3$ can be found as close to the norm of the second largest eigenvalue of $\mathcal{L}$ as possible.
	\end{corollary}
	\vspace{6pt}
	\begin{proof}
		The corollary can be proved with Theorem \ref{thmric}, Theorem \ref{thmcov} and the triangular inequality of 2-norm:
		$$
		\big\|\tilde{P}_{i,k}^{(L)}-P_k\big\|_2\leq \big\|\tilde{P}_{i,k}^{(L)}-P_{i,k}^{(L)}\big\|_2+\big\|P_{i,k}^{(L)}-P_k\big\|_2.
		$$
		For the second part, with the proof of Theorem \ref{thmric} and Theorem \ref{thmcov}, the decay rates $q_1,q_2,q_3$ are actually bounded by that of $Nl_{ij}^{(L)}-1$. Thus, the result follows.
	\end{proof}
	
	\vspace{6pt}
	\begin{remark}
		Corollary \ref{exponential} illustrates that, with the number of fusion step tending to infinity, the SPPS performance of the error covariance matrix $\tilde{P}_{i,k}^{\left(L\right)}$ for each sensor $i$ will exponentially converge to the centralized optimal SPPS performance $P_k$. Thus it reflects the trade-off between estimation accuracy and communication cost among sensor nodes {in the distributed setting, for periodic systems.} In addition, the exponential bound is uniform, namely, it is not related to the time step $k$. The performance gap $\big\|\tilde{P}_{i,k}^{(L)}-P_k\big\|_2$ is separated by the matrix performance $P_{i,k}^{(L)}$. With the Schur-stable {\it monodromy matrix} $\Phi_{P_{i,k}^{(L)}}$, each of $\big\|\tilde{P}_{i,k}^{(L)}-P_{i,k}^{(L)}\big\|_2$ and $\big\|P_{i,k}^{(L)}-P_k\big\|_2$ can be represented with an infinite series form, with which a linear relationship of the perturbation term of the noise matrix $\big\|\big(\tilde{R}_{i,k}^{(L)}\big)^{-1}-R_k^{-1}\big\|_2$ can be obtained. 
	\end{remark}
    \textcolor{black}{\begin{remark}
	Note that due to the possible existence of the high-dimension Jordan block of matrix $\mathcal{L}$, one can not directly state that the inequality $\big\|\big(\tilde{R}_{i,k}^{(L)}\big)^{-1}-R_k^{-1}\big\|_2\leq \tilde{M}\sigma_2^L$ holds, where $\sigma_2$ is the norm of the second largest eigenvalue of $\mathcal{L}$ and $\tilde{M}$ is a scalar not related to $L$. However, one can choose a positive number $\epsilon$ to make the inequality $\big\|\big(\tilde{R}_{i,k}^{(L)}\big)^{-1}-R_k^{-1}\big\|_2\leq \tilde{M}(\sigma_2+\epsilon)^L$ holds, and the $\epsilon$ can be as small as possible, which is the meaning of the ``closeness" in Lemma \ref{lm1} and Corollary \ref{exponential}.
    \end{remark}}
	
	\vspace{6pt}
	%    \textcolor{black}{To summarize, the contents and proof of Theorem \ref{thmcov} are the basis for the derivation of Corollary 1 and 2, and Corollary 2 reveals the connection between the convergence rate $q$ and the spectral property of $\mathcal{L}$. }
	%	, the term $\big(Nl_{ij}^{(L)}-1\big)^2$ is important for analyzing the distance between $P_{i}^{(L)}$ and $\tilde{P}_{i}^{(L)}$. When $\big|Nl_{ij}^{(L)}-1\big|<1$, {it is $\big(Nl_{ij}^{(L)}-1\big)$ that dominates $\big(R_{i}^{(L)}\big)^{-1}\big(R-R_{i}^{(L)}\big)\big(R_{i}^{(L)}\big)^{-1}$}. When $\big|Nl_{ij}^{(L)}-1\big|>1$, it is $\big(Nl_{ij}^{(L)}-1\big)^2$ that dominates $\big(R_{i}^{(L)}\big)^{-1}\big(R-R_{i}^{(L)}\big)\big(R_{i}^{(L)}\big)^{-1}$. 
	
	\section{Simulation}\label{sec4}
	In this section, a numerical experiment on a periodic target system with a periodically scheduled sensor network  is provided to illustrate the effectiveness and correctness of the proposed algorithm. The state transition matrix has the expression
    $$
	\begin{aligned}
	&a_{k}=\begin{pmatrix}
	0.8+0.4*sin(\omega_1 k)&0.5*sin(\omega_2 k)\\
	0.7*cos(\omega_1 k)&0.9+0.3*cos(\omega_2 k)
	\end{pmatrix},\\
	&A_k=\begin{pmatrix}
	a_{k}&0_{2\times 2}\\
	0_{2\times 2}&a_{k}
	\end{pmatrix}.
	\end{aligned}
	$$
	The error covariance matrix $Q$ {takes the form of}
	$$
	\begin{aligned}
	&G=\begin{pmatrix}
	\frac{T^3}{3}&\frac{T^2}{2}\\
	\frac{T^2}{2}&T
	\end{pmatrix}\quad
	&Q=\begin{pmatrix}
	G&0.5G\\
	0.5G&G\\
	\end{pmatrix},
	\end{aligned}
	$$
	where the parameters are set to be $T=1$, $\omega_1=\frac{\pi}{3}$ and $\omega_2=\frac{\pi}{5}$. There are three kinds of sensors with the observation matrix:
	$$
	\begin{aligned}
	&C^{(1)}_{2k-1}=\left[1,0,0,0\right],\quad C^{(1)}_{2k}=\left[0,0,0,0\right],\\
	&C^{(2)}_{2k-1}=\left[0,0,1,0\right],\quad C^{(2)}_{2k}=\left[0,0,0,0\right],\\
	&C^{(3)}_{2k-1}=\left[0,0,0,0\right],\quad C^{(3)}_{2k}=\left[0,0,0,0\right],
	\end{aligned}
	$$
	where $k=1,2,\dots$. Note that sensor $C^{(1)}$ and sensor $C^{(2)}$ are periodically scheduled and the period of the whole system is $\mathcal{T}=30$.
	The whole network consists of 20 sensor nodes, including 3 sensors of kind $C^{(1)}$, 3 sensors of kind $C^{(2)}$, 14 sensors of kind $C^{(3)}$, and $R_{i,k}=1,\;\forall i\in\mathcal{V}$. {The locations of the sensors are randomly set in a $300\times300$ {region} and each sensor is with a communication radius of $130$.} The communication topology of sensor network is randomly generated in the numerical experiments, {as presented in Fig.~\ref{topology}.}
	\begin{figure}
		\centering
		\includegraphics[width=0.4\textwidth]{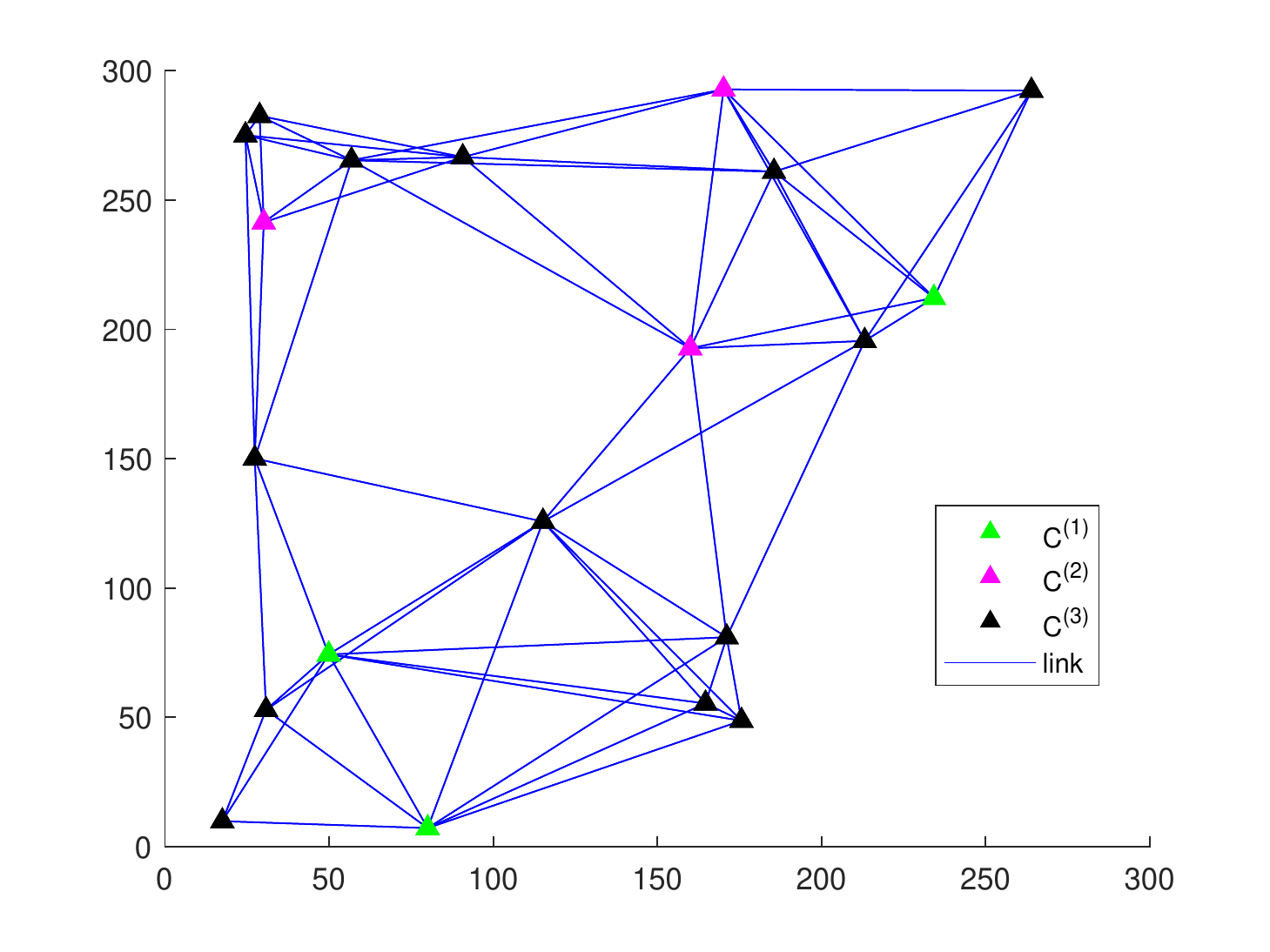}
		\caption{
			Illustration figure for communication topology of the sensor network.}
		\label{topology}	
	\end{figure}
	
	The expression of three kinds of observation matrices indicates that each kind of sensor is able to measure some partial states of the system but none of the pair $\left(A_k,C_{i,k}\right)$ is fully observable. Moreover, the periodically scheduled sensor indicates that the pair $\left(A_k,C_{k}\right)$ may also be unobservable for some $k$. However, the uniform observablity of $\left(A_.,C_{.}\right)$ can be verified and the stability of the algorithm can also be guaranteed for sufficiently large $L$. Thus, the numerical experiments can fully illustrate the effectiveness of the CMDF algorithm on periodic system, {as well as} the relationship between the filtering performance gap and fusion step $L$. Furthermore, the existence of $C^{(3)}$ implies that the CMDF algorithm is tolerant to {the naive nodes}\cite{battistelli2018distributed}.
	
	In the experiment, the SPPS performance of the CMDF algorithm is {evaluated by the mean square error (MSE). Using} {Monte Carlo} method, the filtering process is run 100 steps for each simulation and $h=1500$ times in total. 
	
	\begin{figure}
		\centering
		\includegraphics[width=0.4\textwidth]{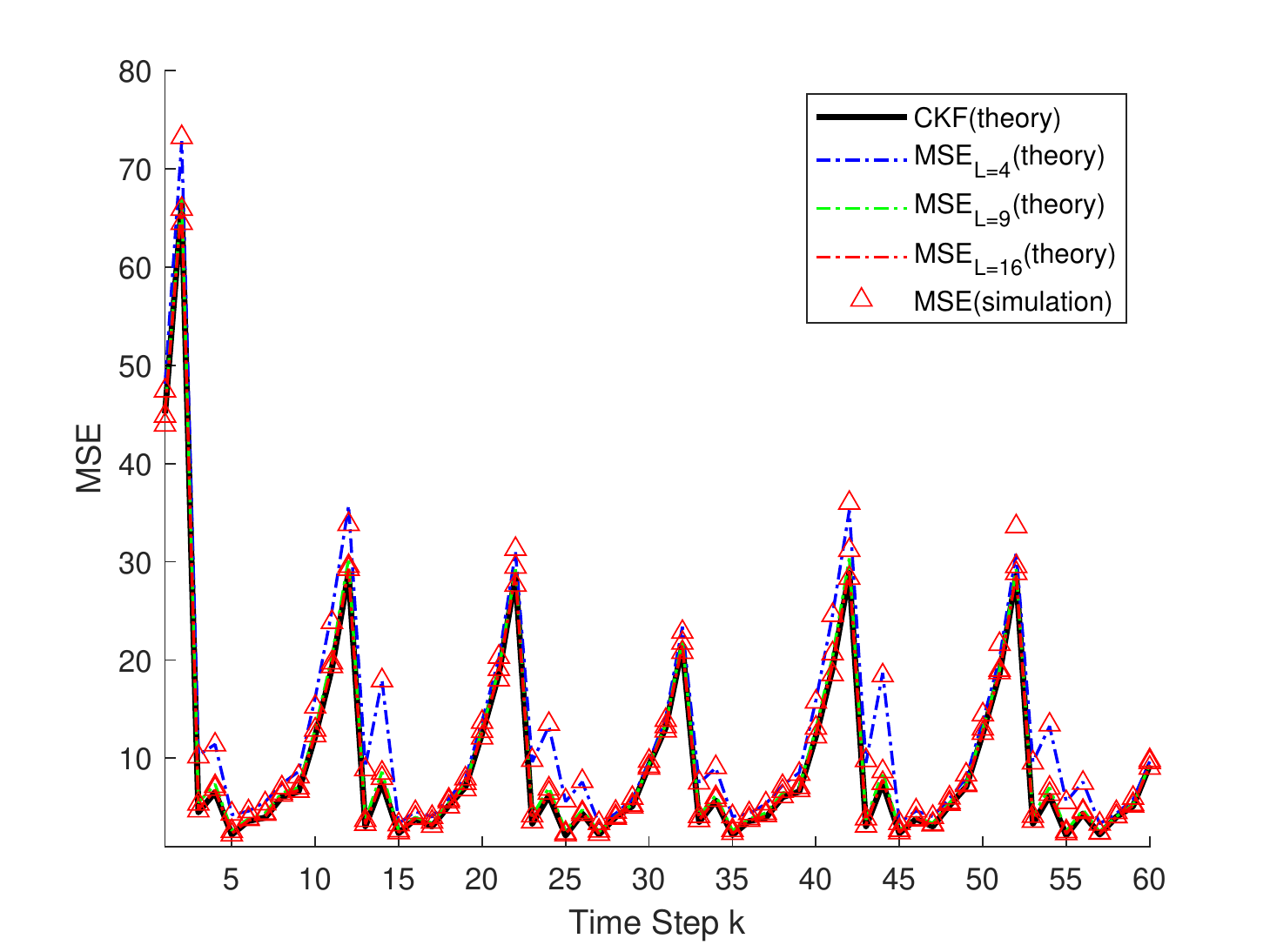}
		\caption{
			Illustration figure for the convergence of the iteration of real covariance $\tilde{P}_{i,k|k-1}$, where CKF (theory) denotes the iteration of the MSE of the centralized Kalman filtering algorithm, $\text{MSE}_{L=i}$ (theory) denotes the theoretical MSE of the $5$-th sensor with the fusion step equal to $i$, and MSE (simulation) denotes the real $\text{MSE}_{5,k}$ calculated through \eqref{singlemse}.}
		\label{singleConvergence}	
	\end{figure}
	In Fig.~\ref{singleConvergence}, the iteration of the theoretical value of MSE, i.e. the trace of matrix $\tilde{P}_{i,k|k-1}$ and real mean square error 
	\begin{equation}\label{singlemse}
	MSE_{i,k}=\frac{1}{h}\sum_{l=1}^{h}\big\|\hat{x}_{i,k}^{\left(l\right)}-x_{k}^{(l)}\big\|_{2}^{2}
	\end{equation}
	are presented under different value of fusion step $L$, where $\hat{x}_{i,k}^{\left(l\right)}$ and $x_{k}^{(l)}$ denote the estimated state and real state in the $l$-th simulation, respectively and the data of the $14$-th sensor is presented. From Fig.~\ref{singleConvergence}, one can obtain that with the increase of time step $k$, the iteration of $\tilde{P}_{i,k|k-1}$ converges to the SPPS solution to the DPLE. Meanwhile, with the increase of $L$, the SPPS solution also converges to the optimal SPPS performance of CKF.
	
	With the result {obtained} in Section \ref{sec3}.B and the result of the above numerical experiment, in each simulation, the performance of the CMDF algorithm converges to the SPPS performance. Thus, the average of the estimated error in one period of the periodic zone {is used} to calculate the average mean square error,  i.e.,
	\begin{equation}\label{mymse}
	MSE_{i}=\frac{1}{h*\mathcal{T}}\sum_{l=1}^{h}\sum_{j=1}^{\mathcal{T}}\big\|\hat{x}_{i,\infty+j}^{\left(l\right)}-x_{\infty+j}^{(l)}\big\|_{2}^{2},
	\end{equation}
	where $\infty$ means {sufficiently large} sampling instant $k$.
	
	The average SPPS performance of each sensor with different values of fusion step $L$ is obtained to illustrate the property of exponential convergence. The performance of CMDF algorithm with different fusion step $L$ is presented in Fig.~\ref{partial}. 
	
	\begin{figure}
		\centering
		\includegraphics[width=0.4\textwidth]{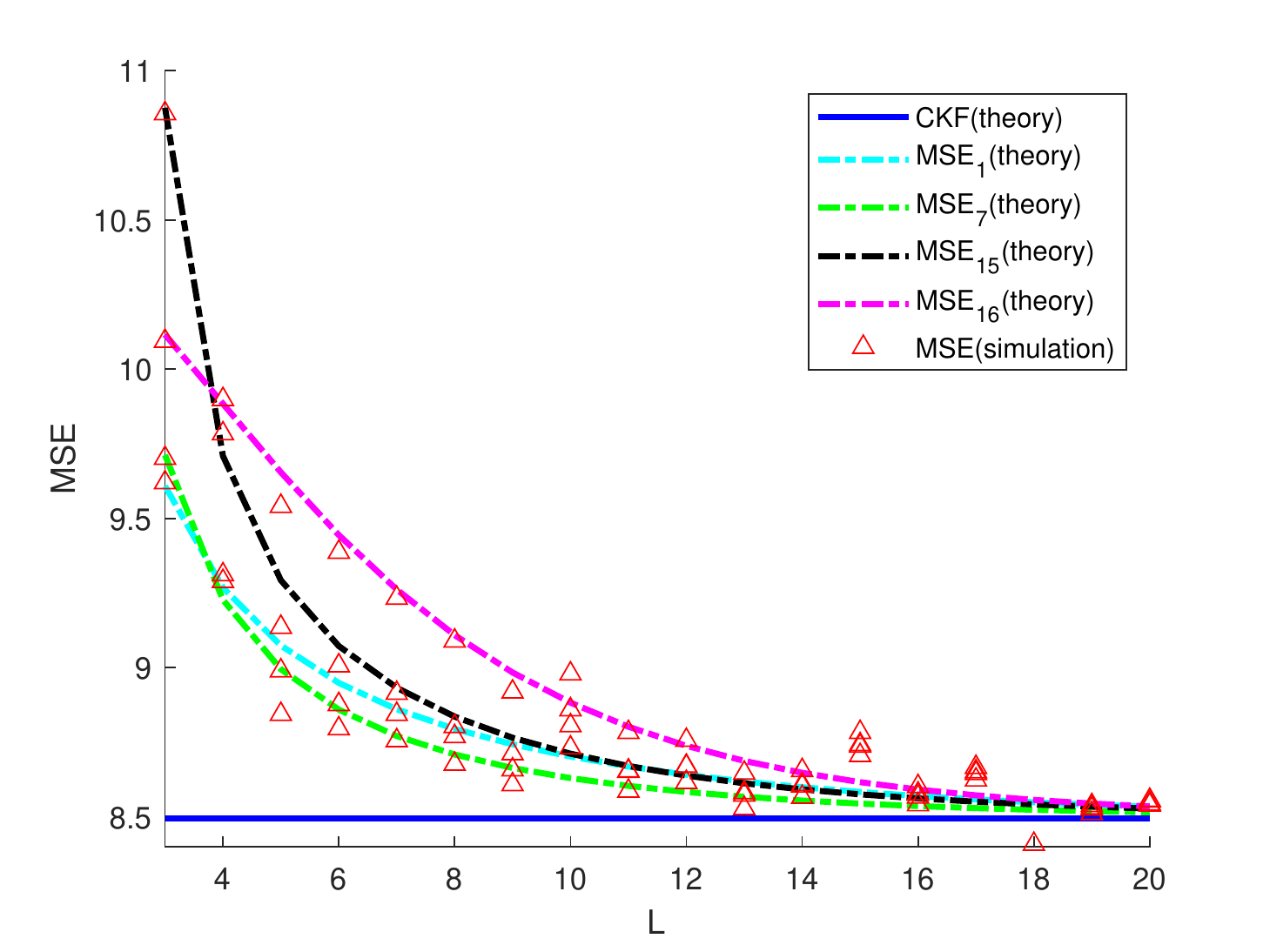}
		\caption{
			Illustration figure for convergence of average SPPS performance with the increasing of $L$, where CKF (theory) denotes the average SPPS MSE of the centralized Kalman filtering algorithm for periodic system, $\text{MSE}_i$ (theory) denotes the average SPPS MSE of the $i$-th sensor calculated through average of \eqref{lyaeq} in one period and MSE (simulation) denotes the average SPPS MSE calculated through \eqref{mymse}.}
		\label{partial}
	\end{figure}
	
	Fig.~\ref{partial} illustrates the SPPS performance of mean square error for partial sensors to fully illustrate the exponential convergence property, which verifies the results of Theorem \ref{thmric}, Theorem \ref{thmcov}.
	
	\textcolor{black}{In addition, to verify the theory obtained in Corrollary \ref{exponential}, the convergence rate $q$ of the performance gap for each sensor node $i$ is investigated. For a prescribed $L$, the average theoretical performances for each sensor $i$ and CKF are respectively calculated as}
	\textcolor{black}{ 
	\begin{equation}\label{averageperform}
			\tilde{P}_{i}^{(L)}=\frac{1}{\mathcal{T}}\sum_{j=1}^{\mathcal{T}}\text{tr}\big(\tilde{P}_{i,\infty+j}^{(L)}\big),\quad P=\frac{1}{\mathcal{T}}\sum_{j=1}^{\mathcal{T}}\text{tr}\big(P_{\infty+j}\big),
	\end{equation}
	where $\text{tr}\left(\cdot\right)$ is the trace of matrix. The convergence rate $q^{(i)}$ for each sensor $i$ is calculated as
	\begin{equation}\label{rate}
	q^{(i)}=\frac{\tilde{P}_{i}^{(L+1)}-P}{\tilde{P}_{i}^{(L)}-P},
	\end{equation}
    which reflects the convergence rate of the performance gap. The results are proposed in Fig.~\ref{Ratio}.
    It is calculated that the norm of the second largest eigenvalue of the doubly stochastic matrix $\mathcal{L}$ is $\sigma=0.92$, and Fig.~\ref{Ratio} depicts that the convergence rate $q^{(i)}$ of the performance gap for each sensor $i$ is smaller than $\sigma$, which verifies Corollary \ref{exponential}.}
	
	\begin{figure}
		\centering
		\includegraphics[width=0.4\textwidth]{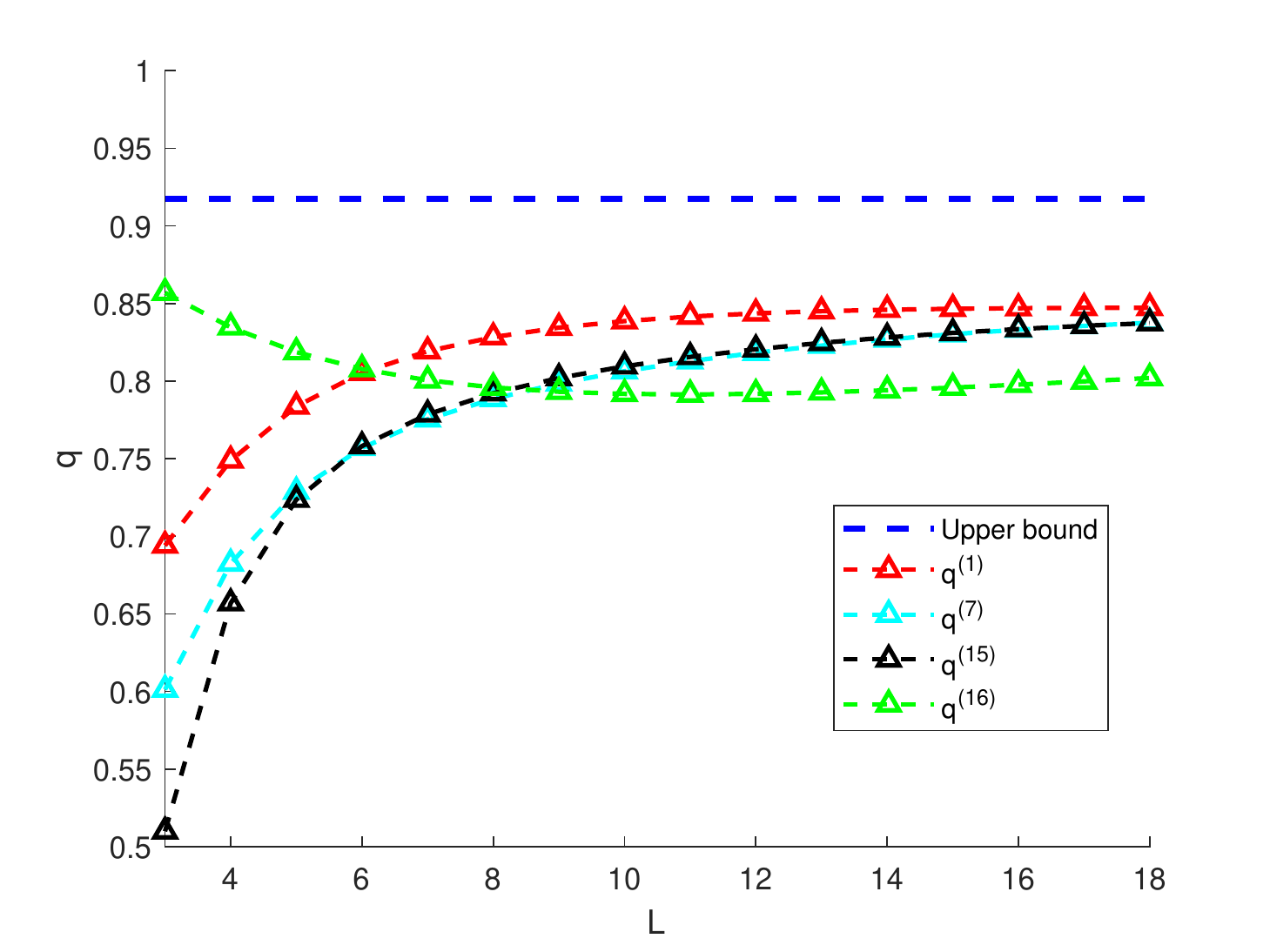}
		\caption{
			Illustration figure for the convergence rate of the performance gap of CMDF and CKF with the increasing of $L$, where Upper bound denotes the norm of the second largest eigenvalue of the doubly stochastic matrix $\mathcal{L}$, $q^{(i)}$ denotes the convergence rate of performance gap calculated through \eqref{rate}.}
		\label{Ratio}
	\end{figure}
    
    \textcolor{black}{Finally, to demonstrate the superiority of CMDF over other distributed filtering algorithms, the comparison of the performance of Algorithm \ref{alg1} proposed in this paper with CIDF proposed in \cite{battistelli2014kullback} is also made, where the performance metric is chosen to be average MSE, calculated through \eqref{mymse}. For different fusion step $L$, the performance of sensor 1 and sensor 16 with CMDF and CIDF are proposed in Fig.~\ref{Comparison}, from which it is revealed that the performance of CMDF is much better than that of CIDF for sufficiently large $L$.} 
    \begin{figure}
 	 \centering
 	 \includegraphics[width=0.4\textwidth]{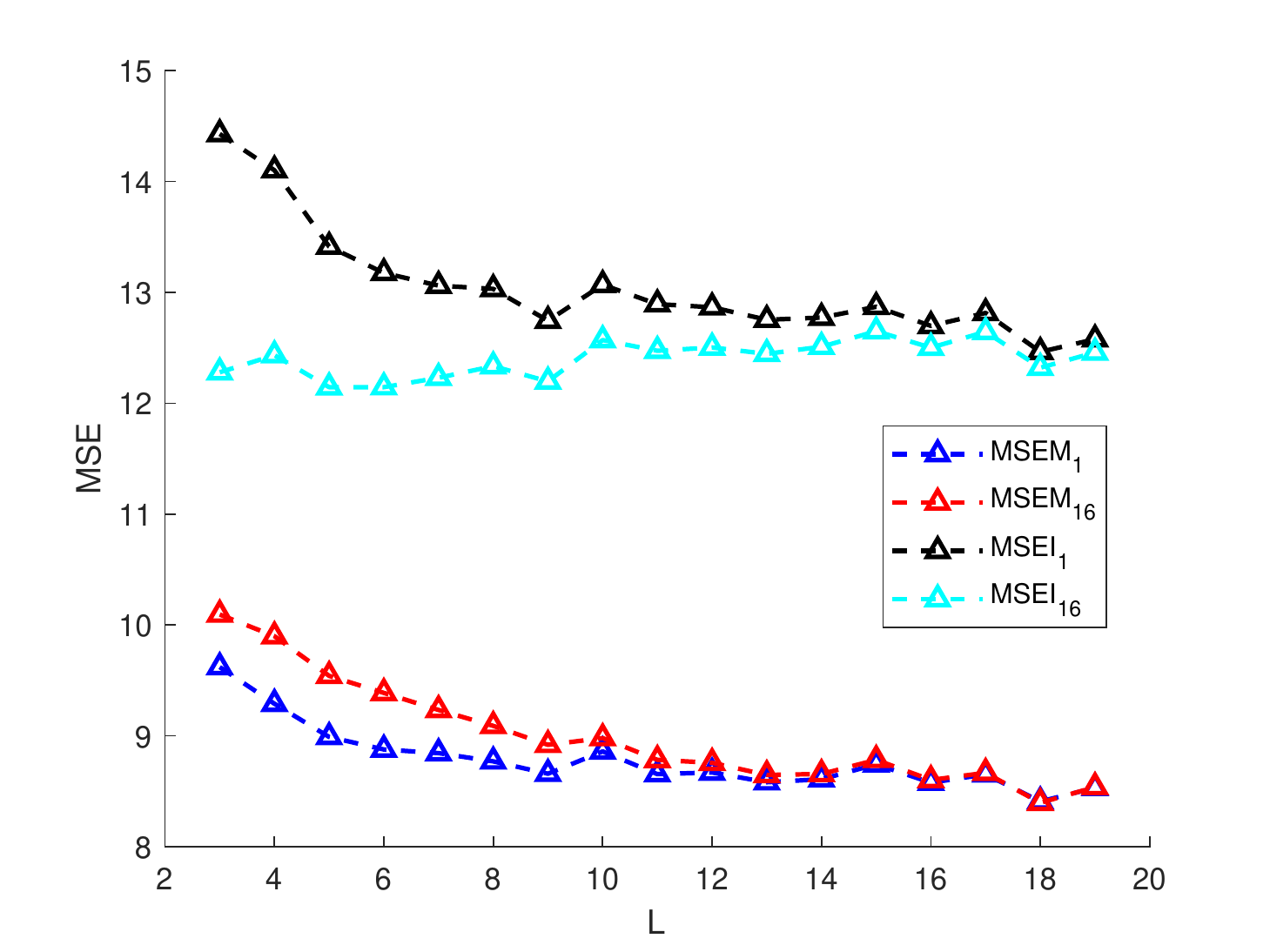}
  	 \caption{
 	 	Illustration figure for the comparison of CMDF and CIDF with the increasing of $L$, where $MSEM_i$ denotes the average MSE of sensor $i$ with CMDF, and $MSEI_{i}$ denotes the average MSE of sensor $i$ with CIDF.}
  	 \label{Comparison}
    \end{figure}
	
	%  and Fig.\ref{total} illustrates the steady-state performance with different $L$ for the whole sensor network. Fig.\ref{partial} and Fig.\ref{total} 

	%\begin{figure}
	%	\centering
	%	\includegraphics[width=0.4\textwidth]{total.eps}
	%	\caption{
	%		Illustration figure for convergence of filtering performance with the increasing of $L$, where $\text{MSE}_{L=k}$(theory) denotes MSE for all sensors with fusion step $L$ equal to $k$.}
	%    \label{total}
	%\end{figure}
	
	%\begin{remark}
	%	It is worth mentioning that the filtering performance of the algorithm in \cite{kamal2013information} and \cite{talebi2019distributed} also showed exponential convergence {to the centralized setting with the increasing of $L$}, in their numerical experiments but without proof. In this paper, a rigorous proof of the exponential convergence of the CMDF algorithm {is provided}.
	%\end{remark}
	
	\section{Conclusion}\label{sec5}
	\textcolor{black}{In this paper, the performance of the CMDF algorithm on periodic system has been investigated. The stability condition of the filtering algorithm has been formulated, and the correponding performance indices have been simplified as solutions to DPRE and DPLE, respectively. Based on the newly discovered properties of solutions to DPRE and DPLE, the trade-off between the fusion step and the SPPS performance of the filtering algorithm {has been revealed}, including the SPPS parameter matrix and SPPS estimation error covariance matrix. As the number of fusion step tends to infinity, both the parameter matrix and error covariance matrix converge to the centralized optimal SPPS ones exponentially.
    Future works include parameter tuning technique for controlling the SPPS performance of the algorithm, or for a better convergence rate of the performance gap.}

    \appendices
	
	%\section{Proof of Lemma \ref{lm3}}\label{Pflm3}
	%	Consider the Schur Decomposition of matrix $A$ as
	%	$
	%	A=U^H TU
	%	$, where $U$ is a unitary matrix and $T$ is an upper-triangular matrix with the diagonal elements being the eigenvalues of $A$. As the singular value is unitary invariant, it is {easy} to obtain  
	%	$$
	%	\lambda_i \left(T\right)=\lambda_i \left(A\right),\qquad \big\|T\big\|_2=\big\|A\big\|_2.
	%	$$ 
	%	By the definition of 2-norm, one has
	%	$$
	%	\big\|A\big\|_2^2=\max\left(\lambda\left(AA^T\right)\right)=\max\left(\lambda\left(TT^T\right)\right).
	%	$$
	%	As the maximum eigenvalue of the positive definite matrix $AA^T$ is larger than all the diagonal elements of $AA^T$, one has
	%	$$
	%	\big\|A\big\|_2^2\ge \max_{i}\sum_{j=1}^n t_{ij}^2,\qquad\max_{i,j}\left|t_{ij}\right|\leq \big\|A\big\|_2,
	%	$$
	%	where $t_{ij}$ is the $\left(i,j\right)$-th element of $T$. Denote $M_T=\max_{i,j}\left|t_{ij}\right|$. {By Corollary 3.15 in \cite{dowler2013bounding}, one has}
	%	$$
	%	\begin{aligned}
	%		\left\|A^k\right\|_2&\leq\sqrt{n}\sum_{j=0}^{n-1}\binom{n-1}{j}\binom{k}{j}M_T^j\rho\left(A\right)^{k-j}\\
	%		&\leq\sqrt{n}\sum_{j=0}^{n-1}\binom{n-1}{j}\binom{k}{j}\big\|A\big\|_2^j\rho\left(A\right)^{k-j}.
	%	\end{aligned}
	%	$$ $\hfill \square$
	
	\section{Proof of Lemma \ref{lmlya}} \label{prfLya}
	\textcolor{black}{First, note that $\bar{A}_.$ is asymptotically stable and $A_.$ converges to $\bar{A}_.$, thus for sufficient large $k_0$, there is that $A_.$ is asymptotically stable for all $k\ge k_0$, i.e., the matrix $\prod_{i=1}^\mathcal{T}A_{k+i}$ is Schur stable $\forall k\ge k_0$. Hence, there exists a matrix $P>0$ such that $P_k\leq P,\;\forall k\in\mathbb{N}$.}
	
	\textcolor{black}{Denote the term $\tilde{A}_k\triangleq A_k-\bar{A}_k$, and $\tilde{Q}_k\triangleq Q_k-\bar{Q}_k$. Then the iteration of $P_{k}-\bar{P}_k$ can be reformulated as
	$$
	\begin{aligned}
	P_{k+1}-\bar{P}_{k+1}&=\bar{A}_k\big(P_k-\bar{P}_k\big)\bar{A}_k^T+\tilde{A}_kP_k\bar{A}_k^T+\bar{A}_kP_k\tilde{A}_k^T\\
	&+\tilde{A}_kP_k\tilde{A}_k^T+\tilde{Q}_k.
	\end{aligned}
	$$
	Performing the above iteration for $\mathcal{T}$ times, one has
	$$
	P_{k+\mathcal{T}}-\bar{P}_{k+\mathcal{T}}=\bar{\Phi}_{k+\mathcal{T},k}\big(P_k-\bar{P}_k\big)\bar{\Phi}_{k+\mathcal{T},k}^T+\Psi_{k+\mathcal{T},k},
	$$
	where
	\begin{equation}\label{asymp}
	\begin{aligned}
	\bar{\Phi}_{k+\mathcal{T},k}=&\prod_{i=k}^{k+\mathcal{T}-1}\bar{A}_i,\quad \bar{\Phi}_{k,k}=I,\\
	\Psi_{k+\mathcal{T},k}=&\sum_{i=k}^{k+\mathcal{T}-1}\bar{\Phi}_{k+\mathcal{T},i+1}\big(\tilde{A}_iP_i\tilde{A}_i^T+\tilde{Q}_i\\
	&+\tilde{A}_iP_i\bar{A}_i^T+\bar{A}_iP_i\tilde{A}_i^T\big)\bar{\Phi}_{k+\mathcal{T},i+1}^T.
	\end{aligned}
	\end{equation}
	Note that $\lim_{k\to\infty}\tilde{A}_k=\textbf{O}$, $\lim_{k\to \infty}\tilde{Q}_k=\textbf{O}$, $\Phi_{k+\mathcal{T},i}$ and $P_i$ are uniformly bounded. Therefore, for any $\epsilon>0$, one can find a sufficient large $k_0^{'}$ such that $\big\|\Psi_{k+\mathcal{T},k}\big\|_2\leq \epsilon,\;\forall k\ge k_0^{'}$. Through performing the iteration \eqref{asymp} for $h$ times, one has
	$$
	\begin{aligned}
	&P_{k+h\mathcal{T}}-\bar{P}_{k+h\mathcal{T}}=\bar{\Phi}_{k+\mathcal{T},k}^h\big(P_k-\bar{P}_k\big)\big(\bar{\Phi}_{k+\mathcal{T},k}^h\big)^T\\
	&\qquad\qquad\qquad+\sum_{i=1}^{h}\bar{\Phi}_{k+\mathcal{T},k}^{h-i}\Psi_{k+i\mathcal{T},k+(i-1)\mathcal{T}}\big(\bar{\Phi}_{k+\mathcal{T},k}^{h-i}\big)^T,\\
	&\big\|P_{k+h\mathcal{T}}-\bar{P}_{k+h\mathcal{T}}\big\|_2\leq \big\|\bar{\Phi}_{k+\mathcal{T},k}^h\big\|_2^2\big\|P_{k}-\bar{P}_{k}\big\|_2\\
	&\qquad\qquad\qquad+\sum_{i=1}^{h}\big\|\bar{\Phi}_{k+\mathcal{T},k}^{h-i}\big\|_2^2\big\|\Psi_{k+i\mathcal{T},k+(i-1)\mathcal{T}}\big\|_2.
	\end{aligned}
	$$
	The above equations holds due to the facts that $\bar{\Phi}_{k+\mathcal{T},k}=\bar{\Phi}_{k+h\mathcal{T},k+(h-1)\mathcal{T}}$ and $\bar{\Phi}_{k+\mathcal{T},k}^h=\bar{\Phi}_{k+h\mathcal{T},k}$.
	For any $k\ge k_0$, one has
	$$
	\begin{aligned}
	\big\|P_{k+h\mathcal{T}}-\bar{P}_{k+h\mathcal{T}}\big\|_2\leq& \big\|\bar{\Phi}_{k+\mathcal{T},k}^h\big\|_2^2\big\|P_{k}-\bar{P}_{k}\big\|_2\\
	&+\epsilon\sum_{i=1}^{h}\big\|\bar{\Phi}_{k+\mathcal{T},k}^{h-i}\big\|_2^2.
	\end{aligned}
	$$
	Due to the asymptotic stability of $\bar{A}_.$, the matrix $\bar{\Phi}_{k+\mathcal{T},k}$ is Schur stable. One has $\lim_{h\to \infty}\big\|\bar{\Phi}_{k+\mathcal{T},k}^h\big\|_2^2=0$ and $\sum_{i=1}^{h}\big\|\bar{\Phi}_{k+\mathcal{T},k}^h\big\|_2^2$ is uniformly bounded for any $h\in\mathbb{N}$. Therefore, for sufficiently large $k$ and $h$, the term $\big\|P_{k+h\mathcal{T}}-\bar{P}_{k+h\mathcal{T}}\big\|_2$ can be arbitrarily small. }
	
	\textcolor{black}{With the above analysis, Lemma \ref{lmlya} is proved.}

	\textcolor{black}{\section{Proof of Lemma \ref{lmmono}} \label{pfmono}
	Consider two matrix sequences with 
	$$
	\begin{aligned}
	\bar{P}_{i,k+1}=&A_k\bar{P}_{i,k}A_k^{T}+Q_k-A_k\bar{P}_{i,k}C_k^T\\
	&\times\Big(C_k \bar{P}_{i,k}C_k^T+R_{i,k}\Big)^{-1}C_k \bar{P}_{i,k}A_k^{T},\; i=1,\,2,
	\end{aligned}
	$$
	the initial values of which are chosen to satisfy the inequality $\bar{P}_{1,0}\ge \bar{P}_{2,0}$. With Lemma \ref{lm2}, one has
	$$
	\begin{aligned}
	\bar{P}_{1,1}=&A_0\left(\bar{P}_{1,0}^{-1}+C_0^TR_{1,0}^{-1}C_0\right)^{-1}A_0^{T}+Q_0\\
	\ge& A_0\left(\bar{P}_{2,0}^{-1}+C_0^TR_{2,0}^{-1}C_0\right)^{-1}A_0^{T}+Q_0=\bar{P}_{2,1},
	\end{aligned}
	$$ 
	which indicates that $\bar{P}_{1,k}\ge \bar{P}_{2,k},\;\forall k\in\mathbb{N}$. In light of Theorem 4 in \cite{bittani1988}, these two matrix sequences will converge to the SPPS solution $P_{1,.}$ and $P_{2,.}$, respectively. Thus this lemma is proved with the order preservation property of limit.}

	\section{Proof of Lemma \ref{lmstable}}\label{Pflmstable}
	\textcolor{black}{Consider the DPRE
	$$
	\begin{aligned}
	P_{k+1}=&A_kP_kA_k^T+Q_k-A_kP_kC_k^T\\
	&\qquad\qquad\times\big(C_kP_kC_k^T+R_k\big)^{-1}C_kP_kA_k^T,
	\end{aligned}
	$$
	which can be rewritten as
	\begin{equation}\label{eqstable}
	P_{k+1} = \tilde{A}_{P_k} P_k\tilde{A}_{P_k}^T+Q_k+K_{P_k}R_kK_{P_k}^T,
	\end{equation}
	where $K_{P_k} = A_kP_kC_k^T\left(C_kP_kC_k^T+R_k \right)^{-1}$. Performing the iteration \eqref{eqstable} for $\mathcal{T}$ times, one has
	\begin{equation}\label{eqmulti}
	\begin{aligned}
	P_{k+\mathcal{T}} =& \Phi_{P_k} P_k\Phi_{P_k}^T
	+\sum_{i=0}^{\mathcal{T}-1}\tilde{\Phi}_{k+\mathcal{T},k+i+1} \\&\times\big(Q_{k+i}+K_{P_{k+i}}R_{k+i}K_{P_{k+i}}^T\big)\tilde{\Phi}_{k+\mathcal{T},k+i+1}^T,
	\end{aligned}
	\end{equation}
	where $\tilde{\Phi}_{k+h,k}=\prod_{i=k}^{k+h-1}\tilde{A}_{P_i},\;\tilde{\Phi}_{k,k}=I$.
	Consider the left eigenvector $x$ of $\Phi_{P_k}$ with corresponding eigenvalue $\lambda$, i.e., $x\Phi_{P_k}=\lambda x$. Pre- and post-multiplying the equation \eqref{eqmulti} with $x$ and $x^H$, respectively, one has
	$$
	\begin{aligned}
	xP_{k+\mathcal{T}}x^H&\ge\left|\lambda\right|^2xP_kx^H+xQ_{k+\mathcal{T}-1}x^H.
	\end{aligned}
	$$
	The above inequality holds due to the fact that the matrix $K_{P_{k+i}}R_{k+i}K_{P_{k+i}}^T$ is positive semi-definite. {Dividing} both side of the above inequality with $xPx^H$, together with the fact that $P_{k+\mathcal{T}}=P_k$, one has
	$$
	\left|\lambda\right|^2\leq 1-\frac{xQ_{k+\mathcal{T}-1}x^H}{xP_kx^H}\leq 1-\frac{\lambda_{min}\left(Q_.\right)}{\lambda_{max}\left(P_.\right)},
	$$
	thus the inequality of the spectral radius is proved. }
	
	\textcolor{black}{For the inequality on the 2-norm, consider the singular value decomposition $\Phi_{P_k}=U\Sigma V^T$, where $\Sigma=\mathrm{diag}\left\lbrace \sigma_1,\dots,\sigma_n\right\rbrace$, $\sigma_1\ge\dots\ge\sigma_n\ge0$. Let $x=\left(1,0,\dots,0\right)U^T$. Pre- and post-multiplying the equation \eqref{eqmulti} with $x$ and $x^T$, respectively, one has
	$$
	\begin{aligned}
	xP_{k+\mathcal{T}}x^T\ge\sigma_1^2\left(V^TP_kV\right)_{1,1},
	\end{aligned}
	$$
	where $\big(V^TP_kV\big)_{1,1}$ is the $(1,1)$-th element of matrix $V^TP_kV$. Note that $\left(V^TP_kV\right)_{1,1}\ge \lambda_{min}\left(V^TP_kV\right)=\lambda_{min}\left(P_k\right)$. Thus, one has
	$$
	\sigma_1^2\leq \frac{\lambda_{max}\left(P_k\right)}{\lambda_{min}\left(P_k\right)}\leq \frac{\lambda_{max}\left(P_.\right)}{\lambda_{min}\left(Q_.\right)},
	$$
	where the last inequality {holds since} $P_{k+1}\ge Q_k$,$\;\forall k\in\mathbb{N}$. $\hfill \square$}

\textcolor{black}{\bibliographystyle{IEEEtran}
\bibliography{ref}}

\begin{thebibliography}{10}
\providecommand{\url}[1]{#1}
\csname url@samestyle\endcsname
\providecommand{\newblock}{\relax}
\providecommand{\bibinfo}[2]{#2}
\providecommand{\BIBentrySTDinterwordspacing}{\spaceskip=0pt\relax}
\providecommand{\BIBentryALTinterwordstretchfactor}{4}
\providecommand{\BIBentryALTinterwordspacing}{\spaceskip=\fontdimen2\font plus
\BIBentryALTinterwordstretchfactor\fontdimen3\font minus
  \fontdimen4\font\relax}
\providecommand{\BIBforeignlanguage}[2]{{%
\expandafter\ifx\csname l@#1\endcsname\relax
\typeout{** WARNING: IEEEtran.bst: No hyphenation pattern has been}%
\typeout{** loaded for the language `#1'. Using the pattern for}%
\typeout{** the default language instead.}%
\else
\language=\csname l@#1\endcsname
\fi
#2}}
\providecommand{\BIBdecl}{\relax}
\BIBdecl

\bibitem{bittani1988}
S.~Bittanti, P.~Colaneri, and G.~De~Nicolao, ``The difference periodic riccati
  equation for the periodic prediction problem,'' \emph{IEEE Transactions on
  Automatic Control}, vol.~33, no.~8, pp. 706--712, 1988.

\bibitem{cole1992nonlinear}
J.~W. Cole and R.~A. Calico, ``Nonlinear oscillations of a controlled periodic
  system,'' \emph{Journal of Guidance, Control, and Dynamics}, vol.~15, no.~3,
  pp. 627--633, 1992.

\bibitem{churilov2012state}
A.~Churilov, A.~Medvedev, and A.~Shepeljavyi, ``A state observer for continuous
  oscillating systems under intrinsic pulse-modulated feedback,''
  \emph{Automatica}, vol.~48, no.~6, pp. 1117--1122, 2012.

\bibitem{nie2011optimal}
J.~Nie, E.~Sheh, and R.~Horowitz, ``Optimal {$H_{\infty}$} control for hard
  disk drives with an irregular sampling rate,'' in \emph{Proceedings of the
  American Control Conference}.\hskip 1em plus 0.5em minus 0.4em\relax IEEE,
  2011, pp. 5382--5387.

\bibitem{meyer1975unified}
R.~Meyer and C.~Burrus, ``A unified analysis of multirate and periodically
  time-varying digital filters,'' \emph{IEEE Transactions on Circuits and
  Systems}, vol.~22, no.~3, pp. 162--168, 1975.

\bibitem{shi2011optimal}
L.~Shi, P.~Cheng, and J.~Chen, ``Optimal periodic sensor scheduling with
  limited resources,'' \emph{IEEE Transactions on Automatic Control}, vol.~56,
  no.~9, pp. 2190--2195, 2011.

\bibitem{xie1991h}
L.~Xie, C.~E. de~Souza, and M.~Fragoso, ``{$H_{\infty}$} filtering for linear
  periodic systems with parameter uncertainty,'' \emph{Systems \& Control
  Letters}, vol.~17, no.~5, pp. 343--350, 1991.

\bibitem{soderstrom2005periodic}
T.~S{\"o}derstr{\"o}m, T.~Wigren, and E.~Abd-Elrady, ``Periodic signal analysis
  by maximum likelihood modeling of orbits of nonlinear odes,''
  \emph{Automatica}, vol.~41, no.~5, pp. 793--805, 2005.

\bibitem{dragan2012optimal}
V.~Dragan, ``Optimal filtering for discrete-time linear systems with
  multiplicative white noise perturbations and periodic coefficients,''
  \emph{IEEE Transactions on Automatic Control}, vol.~58, no.~4, pp.
  1029--1034, 2012.

\bibitem{shi2013approximate}
D.~Shi and T.~Chen, ``Approximate optimal periodic scheduling of multiple
  sensors with constraints,'' \emph{Automatica}, vol.~49, no.~4, pp. 993--1000,
  2013.

\bibitem{orihuela2014periodicity}
L.~Orihuela, A.~Barreiro, F.~G{\'o}mez-Estern, and F.~R. Rubio, ``Periodicity
  of {Kalman}-based scheduled filters,'' \emph{Automatica}, vol.~50, no.~10,
  pp. 2672--2676, 2014.

\bibitem{olfati2007distributed}
R.~Olfati-Saber, ``Distributed {Kalman} filtering for sensor networks,'' in
  \emph{Proceedings of the IEEE Conference on Decision and Control}, 2007, pp.
  5492--5498.

\bibitem{olfati2009kalman}
------, ``Kalman-consensus filter: Optimality, stability, and performance,'' in
  \emph{Proceedings of the IEEE Conference on Decision and Control held jointly
  with 28th Chinese Control Conference}, 2009, pp. 7036--7042.

\bibitem{cattivelli2010diffusion}
F.~S. Cattivelli and A.~H. Sayed, ``Diffusion strategies for distributed
  {Kalman} filtering and smoothing,'' \emph{IEEE Transactions on Automatic
  Control}, vol.~55, no.~9, pp. 2069--2084, 2010.

\bibitem{kamal2013information}
A.~T. Kamal, J.~A. Farrell, and A.~K. Roy-Chowdhury, ``Information weighted
  consensus filters and their application in distributed camera networks,''
  \emph{IEEE Transactions on Automatic Control}, vol.~58, no.~12, pp.
  3112--3125, 2013.

\bibitem{battistelli2014kullback}
G.~Battistelli and L.~Chisci, ``Kullback--{Leibler} average, consensus on
  probability densities, and distributed state estimation with guaranteed
  stability,'' \emph{Automatica}, vol.~50, no.~3, pp. 707--718, 2014.

\bibitem{Battistelli2015Linear}
G.~{Battistelli}, L.~{Chisci}, G.~{Mugnai}, A.~{Farina}, and A.~{Graziano},
  ``Consensus-based linear and nonlinear filtering,'' \emph{IEEE Transactions
  on Automatic Control}, vol.~60, no.~5, pp. 1410--1415, 2015.

\bibitem{Wang20181300}
S.~Wang and W.~Ren, ``On the convergence conditions of distributed dynamic
  state estimation using sensor networks: A unified framework,'' \emph{IEEE
  Transactions on Control Systems Technology}, vol.~26, no.~4, pp. 1300--1316,
  2018.

\bibitem{he2020distributed}
X.~He, W.~Xue, X.~Zhang, and H.~Fang, ``Distributed filtering for uncertain
  systems under switching sensor networks and quantized communications,''
  \emph{Automatica}, vol. 114, p. 108842, 2020.

\bibitem{duan2020distributed}
P.~Duan, Z.~Duan, G.~Chen, and L.~Shi, ``Distributed state estimation for
  uncertain linear systems: A regularized least-squares approach,''
  \emph{Automatica}, vol. 117, p. 109007, 2020.

\bibitem{li2020Bound}
W.~{Li}, Z.~{Wang}, D.~W.~C. {Ho}, and G.~{Wei}, ``On boundedness of error
  covariances for {Kalman} consensus filtering problems,'' \emph{IEEE
  Transactions on Automatic Control}, vol.~65, no.~6, pp. 2654--2661, 2020.

\bibitem{Sayed20219353995}
S.~P. Talebi, S.~Werner, V.~Gupta, and Y.-F. Huang, ``On stability and
  convergence of distributed filters,'' \emph{IEEE Signal Processing Letters},
  vol.~28, pp. 494--498, 2021.

\bibitem{Wu201863}
Z.~Wu, M.~Fu, Y.~Xu, and R.~Lu, ``A distributed {Kalman} filtering algorithm
  with fast finite-time convergence for sensor networks,'' \emph{Automatica},
  vol.~95, pp. 63--72, 2018.

\bibitem{chenb9416784}
B.~Chen, G.~Hu, D.~W. Ho, and L.~Yu, ``Distributed estimation and control for
  discrete time-varying interconnected systems,'' \emph{IEEE Transactions on
  Automatic Control}, vol.~67, no.~5, pp. 2192--2207, 2022.

\bibitem{zhang2016event}
H.~Zhang, Q.~Hong, H.~Yan, F.~Yang, and G.~Guo, ``Event-based distributed
  {$H_{\infty}$} filtering networks of 2-dof quarter-car suspension systems,''
  \emph{IEEE Transactions on Industrial Informatics}, vol.~13, no.~1, pp.
  312--321, 2016.

\bibitem{shen2019distributed}
B.~Shen, Z.~Wang, D.~Wang, and H.~Liu, ``Distributed state-saturated recursive
  filtering over sensor networks under round-robin protocol,'' \emph{IEEE
  Transactions on Cybernetics}, vol.~50, no.~8, pp. 3605--3615, 2019.

\bibitem{wanrb8401335}
X.~Wan, Z.~Wang, M.~Wu, and X.~Liu, ``{$H_{\infty}$} state estimation for
  discrete-time nonlinear singularly perturbed complex networks under the
  round-robin protocol,'' \emph{IEEE Transactions on Neural Networks and
  Learning Systems}, vol.~30, no.~2, pp. 415--426, 2019.

\bibitem{li2017distributed}
J.-Y. Li, R.~Lu, Y.~Xu, H.~Peng, and H.-X. Rao, ``Distributed state estimation
  for periodic systems with sensor nonlinearities and successive packet
  dropouts,'' \emph{Neurocomputing}, vol. 237, pp. 50--58, 2017.

\bibitem{lijun8651311}
J.-Y. Li, B.~Zhang, R.~Lu, Y.~Xu, and T.~Huang, ``Distributed {$H_{\infty}$}
  state estimator design for time-delay periodic systems over scheduling sensor
  networks,'' \emph{IEEE Transactions on Cybernetics}, vol.~51, no.~1, pp.
  462--472, 2021.

\bibitem{LiuRoundRobin8605375}
S.~Liu, Z.~Wang, G.~Wei, and M.~Li, ``Distributed set-membership filtering for
  multirate systems under the round-robin scheduling over sensor networks,''
  \emph{IEEE Transactions on Cybernetics}, vol.~50, no.~5, pp. 1910--1920,
  2020.

\bibitem{talebi2019distributed}
S.~P. Talebi and S.~Werner, ``Distributed {Kalman} filtering and control
  through embedded average consensus information fusion,'' \emph{IEEE
  Transactions on Automatic Control}, vol.~64, no.~10, pp. 4396--4403, 2019.

\bibitem{liCMweight}
W.~{Li}, G.~{Wei}, D.~W.~C. {Ho}, and D.~{Ding}, ``A weightedly uniform
  detectability for sensor networks,'' \emph{IEEE Transactions on Neural
  Networks and Learning Systems}, vol.~29, no.~11, pp. 5790--5796, 2018.

\bibitem{qian2021consensusbased}
J.~Qian, P.~Duan, Z.~Duan, G.~Chen, and L.~Shi, ``Consensus-based distributed
  filtering with fusion step analysis,'' \emph{Automatica}, vol. 142, p.
  110408, 2022.

\bibitem{duan229750909}
P.~Duan, J.~Qian, Q.~Wang, Z.~Duan, and L.~Shi, ``Distributed state estimation
  for continuous-time linear systems with correlated measurement noise,''
  \emph{IEEE Transactions on Automatic Control}, pp. 1--1, 2022.

\bibitem{GHARESIFARD2012539}
B.~Gharesifard and J.~Cortes, ``Distributed strategies for generating
  weight-balanced and doubly stochastic digraphs,'' \emph{European Journal of
  Control}, vol.~18, no.~6, pp. 539--557, 2012.

\bibitem{Kamagarpour4738989}
M.~{Kamgarpour} and C.~J. {Tomlin}, ``Convergence properties of a decentralized
  {Kalman} filter,'' in \emph{47th IEEE Conference on Decision and Control},
  2008, pp. 3205--3210.

\bibitem{BATTILOTTI2021109589}
S.~Battilotti, F.~Cacace, and M.~d’Angelo, ``A stability with optimality
  analysis of consensus-based distributed filters for discrete-time linear
  systems,'' \emph{Automatica}, vol. 129, p. 109589, 2021.

\bibitem{guttman1946enlargement}
L.~Guttman, ``Enlargement methods for computing the inverse matrix,'' \emph{The
  Annals of Mathematical Statistics}, pp. 336--343, 1946.

\bibitem{bittanti1990algebraic}
S.~Bittanti, P.~Colaneri, and G.~De~Nicolao, ``An algebraic {Riccati} equation
  for the discrete-time periodic prediction problem,'' \emph{Systems \& Control
  Letters}, vol.~14, no.~1, pp. 71--78, 1990.

\bibitem{bittanti1985discrete}
S.~Bittanti and P.~Bolzern, ``Discrete-time linear periodic systems: Gramian
  and modal criteria for reachability and controllability,''
  \emph{International Journal of Control}, vol.~41, no.~4, pp. 909--928, 1985.

\bibitem{battistelli2018distributed}
G.~Battistelli, L.~Chisci, and D.~Selvi, ``A distributed {Kalman} filter with
  event-triggered communication and guaranteed stability,'' \emph{Automatica},
  vol.~93, pp. 75--82, 2018.

\end{thebibliography}

\end{document}